\documentclass[aps, prd, 10pt, notitlepage, superscriptaddress, nofootinbib,numbers,twocolumn]{revtex4-2}

\usepackage[utf8]{inputenc}
\usepackage[T1]{fontenc}
\usepackage{anyfontsize}

\usepackage{amsmath}
\usepackage{amssymb}
\usepackage{amsfonts}
\usepackage{mathrsfs}

\usepackage{microtype}

\usepackage{lipsum,graphicx,afterpage}
\usepackage{epsfig}

\usepackage{bm}

\usepackage{dcolumn}

\usepackage{longtable}
\usepackage{}

\usepackage[dvipsnames]{xcolor}
\usepackage{hyperref}
\hypersetup{
    colorlinks=true,
    citecolor=Purple,
    linkcolor=Purple,
    urlcolor=Purple,
    linktocpage=true,
    breaklinks=true
}

\usepackage[capitalize]{cleveref}

\begin{document}

\title{Tidal forces around the Letelier-Alencar cloud of strings black hole}
\author{Marcos V. de S. Silva}
\email{marcos.sousa@uva.es}
\affiliation{Department of Theoretical Physics, Atomic and Optics, Campus Miguel Delibes, \\ University of Valladolid UVA, Paseo Bel\'en, 7,
47011 - Valladolid, Spain}
\author{T. M. Crispim}
	\email{tiago.crispim@fisica.ufc.br}
	\affiliation{Departamento de F\'isica, Universidade Federal do Cear\'a, Caixa Postal 6030, Campus do Pici, 60455-760 Fortaleza, Cear\'a, Brazil.}
\author{R. R. Landim}
\email{renan@fisica.ufc.br}
\affiliation{Departamento de F\'isica, Universidade Federal do Cear\'a, Caixa Postal 6030, Campus do Pici, 60455-760 Fortaleza, Cear\'a, Brazil.}

\author{Gonzalo J. Olmo}
\email{gonzalo.olmo@uv.es}
\affiliation{Instituto de Física Corpuscular (IFIC), CSIC‐Universitat de Val\`encia, Spain}
\affiliation{Departamento de F\'isica, Universidade Federal do Cear\'a, Caixa Postal 6030, Campus do Pici, 60455-760 Fortaleza, Cear\'a, Brazil.}

\author{Diego S\'aez-Chill\'on G\'omez}
\email{diego.saez@uva.es} 
\affiliation{Department of Theoretical Physics, Atomic and Optics, and Laboratory for Disruptive Interdisciplinary Science (LaDIS), Campus Miguel Delibes, \\ University of Valladolid UVA, Paseo Bel\'en, 7,
47011 - Valladolid, Spain}
\affiliation{Departamento de F\'isica, Universidade Federal do Cear\'a, Caixa Postal 6030, Campus do Pici, 60455-760 Fortaleza, Cear\'a, Brazil.}

\date{\today}

\begin{abstract}
In this work, we investigate relativistic tidal forces around a black hole sourced by a cloud of strings, described by the generalized Letelier-Alencar solution. We first review the original Letelier spacetime and its recent generalization, computing the Kretschmann scalar and showing that the generalized model exhibits a stronger curvature divergence at $r \to 0$ than both Letelier and Schwarzschild cases. We then analyze geodesic motion in this background. For massless particles, we focus on circular photon orbits, while for massive particles, we consider both radial infall and circular motion. We find that the radii of the photon sphere and of the innermost stable circular orbit increase with the cloud of strings parameter $g_s$ and decrease with the length scale $l_s$, and circular orbits cease to exist in certain regions of the parameter space. For radial motion, we compute the radial acceleration and the corresponding tidal forces. In this case, we show that an inversion between stretching and compression may occur, although this regime is typically hidden inside the event horizon. Once the tidal forces are known, we computed the behavior of the displacement vector in order to verify whether the usual stretching behavior induced by tidal forces is preserved. Finally, we study tidal forces for observers in circular motion, showing that the cloud of strings modifies the Keplerian frequency and the tidal force profile even at large distances, and that in this case there is no sign change of the tidal components.
\end{abstract}


\maketitle

\section{Introduction}
Black holes (BHs) are one of the most important solutions of general relativity (GR). Under broad physical conditions, the singularity theorems show that gravitational collapse can lead to spacetime singularities and event horizons, where geodesics cannot be extended and the classical description breaks down \cite{Penrose:1964wq,Penrose:1969pc,Hawking:1970zqf}. For this reason, BH spacetimes are useful both to test GR in the strong field regime and to study open problems such as the nature of singularities, BH thermodynamics, and the information loss problem \cite{Berti:2015itd,Berti:2025hly}.

In the last years, observations have given strong evidence that BHs exist in the Universe. The detections of gravitational waves from binary BH mergers by the LIGO--Virgo--KAGRA Collaboration opened a new way to study strong gravity and to measure the properties of the final BH \cite{LIGOScientific:2016aoc,LIGOScientific:2016sjg,LIGOScientific:2017ycc,LIGOScientific:2020zkf}. In addition, very-long-baseline interferometry observations by the Event Horizon Telescope have produced images of the shadows of the supermassive BHs in M87* and Sgr~A*, probing the geometry close to the event horizon \cite{EventHorizonTelescope:2019dse,EventHorizonTelescope:2019ggy,EventHorizonTelescope:2022wkp}. These results motivate detailed theoretical studies of these spacetimes and the physical effects in their neighborhood \cite{Falcke:1999pj,Vagnozzi:2022moj}. More recently, very heavy binary mergers such as the GW231123 event, involving BHs with a total mass of a few $10^2 M_\odot$, have further extended the mass range over which strong gravity effects can be probed \cite{LIGOScientific:2025rsn,LIGOScientific:2025rid}.

Astrophysical BHs are not isolated systems, but are typically surrounded by matter and fields \cite{Narayan:2005ie}. Gas and plasma in accretion disks, magnetic fields, dark matter halos, cosmological backgrounds, and scalar fields are some examples of environments that can interact with the central object \cite{Cardoso:2021wlq,Konoplya:2021ube,Medved:2003rga,Cunha:2015yba,Herdeiro:2015waa,Herdeiro:2014goa,Rahmatov:2025gpk,Hoshimov:2025hyt,Bakhodirov:2025tpw,Chan:2025oux}. In this situation, one often speaks of ``dirty'' BHs, where the vacuum geometry is modified by an effective stress-energy tensor that represents the surrounding environment \cite{Visser:1992qh,Medved:2003rga}. Such configurations can shift key geometric scales, like characteristic radii associated with orbits and light propagation \cite{Macedo:2015ikq}. Depending on the type and distribution of this ``dirt'', the physical properties of the spacetime are modified in a way that can, in principle, be constrained by observations \cite{Macedo:2024qky,Barausse:2014tra}.

One of the possible types of ``dirt'' around a BH is a cloud of strings. In GR, a cloud of strings can be described as a continuous distribution of one-dimensional objects whose world-sheets are averaged into an effective stress-energy tensor. The first BH solution with a cloud of strings was proposed by Letelier in 1979 \cite{Letelier:1979ej}. In his work, Letelier formulated this description and obtained a spherically symmetric solution surrounded by a cloud of strings, characterized by an anisotropic stress-energy tensor with $T^{t}{}_{t}=T^{r}{}_{r}\propto 1/r^{2}$ and vanishing angular components, which leads to a metric function of the form $f(r)=1-\alpha-2M/r$, where $\alpha$ measures the string density and $M$ is the BH mass \cite{Letelier:1979ej}. This geometry reduces to the Schwarzschild solution when $\alpha=0$ and to a pure cloud of strings when $M=0$. Since then, the Letelier solution and its extensions have been studied in several contexts, including higher-dimensional generalizations \cite{Ghosh:2014pga,deMToledo:2018tjq,Ghosh:2014dqa,Waseem:2023ejh,Sadeghi:2020bsa}, configurations with quintessence or other dark energy components \cite{Mustafa:2022xod,Toledo:2018hav,Atamurotov:2022nim}, and setups with additional matter sources, such as nonlinear electrodynamics \cite{Rodrigues:2022zph,Muniz:2025ugk}. These works have analyzed the horizon structure \cite{Rincon:2018ktz}, geodesic motion \cite{Li:2020zxi,Belhaj:2022kek,He:2021aeo}, thermodynamics, and phase transitions \cite{Singh:2020nwo,Toledo:2019amt,Rodrigues:2022rfj,Sadeghi:2023tzf} in BHs with cloud of strings, as well as their behavior in asymptotically flat and (A)dS backgrounds \cite{Sadeghi:2020ciy}. In this way, the cloud of strings framework provides a simple but nontrivial model of anisotropic matter around BHs, and it can modify physical properties of spacetime in a way that may, in principle, be constrained by current or future observations \cite{Vagnozzi:2022moj}. A more general cloud of strings configuration was recently proposed by Alencar and collaborators \cite{Alencar:2025zyl}, who introduced additional parameters in the metric to modify the contribution of the cloud while preserving spherical symmetry and anisotropy. This Letelier–Alencar solution changes the horizon structure and other properties compared to the Schwarzschild and original Letelier cases.

It is very important to understand how we can extract information from the spacetime geometry in order to distinguish between different BH models. Tidal forces provide one such tool, since they encode how the gravitational field varies in space and appear directly in the geodesic deviation equation \cite{PoissonWill:2014}. They determine how extended bodies are stretched or compressed as they fall into or orbit around a compact object and are closely related to phenomena such as tidal disruption events and to characteristic scales like the Roche limit \cite{hobson2006}. In BH spacetimes, the tidal field influences the motion and internal stresses of stars, compact remnants, and fluid elements in accretion flows and can leave signatures in both electromagnetic and gravitational-wave signals \cite{Komossa:2015qya}. For these reasons, a detailed study of tidal forces in different BH backgrounds can help identify how modifications of the geometry can affect the observable features of surrounding matter.

For the Schwarzschild case, the tidal forces diverge at singularity and never change sign, yielding a persistent stretching behavior \cite{hobson2006,dinverno1992}. For more complex models, such as the Reissner-Nordström BH, a sign reversal of the tidal forces can occur, so that stretching turns into compression, which may or may not be hidden inside the event horizon \cite{Sharif:2018gzj,Crispino:2016pnv,LimaJunior:2022gko}. In the Reissner-Nordström case, the tidal forces also diverge at the origin, but there exists a turning point such that massive bodies will not encounter the singularity \cite{Crispino:2016pnv}. For rotating BHs, the tidal forces do not necessarily diverge as $r \to 0$, but rather at the ring singularity that such solutions can possess \cite{LimaJunior:2020fhs}. In regular spacetimes, that is, those without interior singularities, such as regular BHs or black bounces, the tidal forces do not diverge \cite{Uniyal:2025sdr, Arora:2023ltv,Lima:2020wcb,Sharif:2018gaj,Crispim:2025cql}. However, there are spacetimes in which the tidal forces do not necessarily diverge at the singularity but rather at the event horizon itself \cite{Nandi:2000gt}. Tidal forces have also been studied in several other settings, such as (A)dS spacetimes \cite{Balivada:2023akk,Vandeev:2021yan}, naked singularities \cite{Viththani:2024fod}, alternative theories of gravity \cite{delaCruz-Dombriz:2013gfa,Harko:2012ve,Barausse:2007ys,Arora:2023ijd}, exotic fluids \cite{Uniyal:2022ouc,Shahzad:2017vwi}, and other cases \cite{Turimov:2025qkz,Cordeiro:2025cfo,Javed:2024jad,Albacete:2024qja,Triantafyllopoulos:2024xct,Toshmatov:2023anz,Hammad:2021zxs,Asgher:2023slh,Vieira:2025vwe,Joshi:2024djy,Mushtaq:2025shw,Idrissov:2025ugs,Camilloni:2023rra,Grilli:2024fds}. Recently, with the aim of studying the regularity of certain classes of spacetimes, several works have studied tidal forces in generic spacetimes and identified the criteria that spacetimes must satisfy for the tidal forces not to diverge \cite{LimaJunior:2025uyj,Zhang:2025nim,Magalhaes:2024smm}.

In this work, we focus on tidal forces in the Letelier–Alencar BH with a cloud of strings, in order to assess how this type of ``dir'' modifies the tidal forces around the BH. To this end, we consider both an observer in radial infall into the BH and an observer in circular motion around it.

This work is organized as follows. In Section \ref{SEC:spacetime}, we provide a brief review of the Letelier solution and discuss how Alencar and collaborators generalized it to obtain a new cloud of string solution. In Section~\ref{SEC:geodesics}, we carry out a detailed analysis of null and timelike geodesic motion in this spacetime, examining their main characteristics, such as the photon unstable orbit, the trajectories of test particles, and how these quantities are affected by the string background. In Section~\ref{SEC:TF}, starting from the geodesic deviation equation, we examine the effects of the string medium on the tidal forces experienced by a body in free fall toward the BH, as well as by a body in circular motion around it. Finally, in Section~\ref{SEC:conclusions} we present our concluding remarks.

Throughout this article, we adopt the metric signature $(-,+,+,+)$ and use natural units in which $8\pi G = 1$.

\section{spacetimes}\label{SEC:spacetime}
In this work, we investigate how the presence of a cloud of strings modifies the tidal forces in the vicinity of BHs. The spacetime considered here can be characterized by the following line element:
\begin{equation}
  \label{metricageral}  ds^2 = -f(r)dt^2 + \frac{dr^2}{f(r)} + r^2d\Omega_2^2,
\end{equation}
where $d\Omega_2^2=d\theta^2+\sin^2\theta d\varphi^2$. Although the models considered here can be described by the line element above, 
there are other BH models with a cloud of strings that require more general line elements to be properly described \cite{Rodrigues:2022rfj}.
 
In this work, we will consider two models with a cloud of strings. 
The first is the Letelier solution, which corresponds to the first BH model with a cloud of strings proposed in the literature. The second is a generalization of this solution, recently proposed by Alencar and collaborators. In the next subsections, we provide a more detailed discussion of these models.

\subsection{Letelier spacetime}
One of the simplest and most well-known models that describe the interaction between BHs and an anisotropic distribution of strings is the solution proposed by Letelier \cite{Letelier:1979ej}. In this scenario, the matter source is a cloud of strings.

In order to find a BH solution with a cloud of string, Letelier considered the Nambu-Goto action, given by
\begin{equation}
S_{NG} = \int\mathcal{M} \sqrt{-\gamma}\, d\lambda^0 d\lambda^1,
\label{eq:nambu_goto_action}
\end{equation}
where $\mathcal{M}$ is a dimensionless constant that characterizes the string, $\lambda^A$ ($A = 0, 1$) are timelike and spacelike parameters, $\gamma = \det(\gamma_{AB})$, and $\gamma_{AB}$ is the induced metric on the world-sheet, given by
\begin{equation}
\gamma_{AB} = g_{\mu\nu}(x) \, \frac{\partial x^\mu}{\partial \lambda^A} \, \frac{\partial x^\nu}{\partial \lambda^B}.
\label{eq:induced_metric}
\end{equation}
Here, $g_{\mu\nu}$ is the spacetime metric and $x^\mu(\lambda^A)$ describes the embedding of the world-sheet $\Sigma$ in the spacetime manifold.

For a cloud of strings, one considers a macroscopic averaging of many strings distributed in space, leading to an effective stress-energy tensor. Introducing the antisymmetric bivector
\begin{equation}
\Sigma^{\mu\nu} = \epsilon^{AB} \frac{\partial x^\mu}{\partial \lambda^A} \frac{\partial x^\nu}{\partial \lambda^B},
\label{eq:sigma_bivector}
\end{equation}
with $\epsilon^{AB}$ being the two-dimensional Levi-Civita symbol ($\epsilon^{01} = -\epsilon^{10} = 1$), the Nambu-Goto action is rewritten as
\begin{equation}
    S_{NG}=\int \mathcal{M} \left( -\frac{1}{2}\Sigma^{\mu\nu}\Sigma_{\mu\nu}\right)^{-1/2}d\lambda^0d\lambda^1,
\end{equation}
and, the stress-energy tensor for a cloud of strings is then given by
\begin{equation}
T^{\mu\nu} = \rho \, \frac{\Sigma^{\mu}{}_{\lambda} \Sigma^{\lambda\nu}}{\sqrt{-\gamma}},
\label{eq:Tmunu_string}
\end{equation}
where $\rho$ is the proper density that characterized the cloud of strings.

From the conservation of the stress-energy tensor, $\nabla_\mu T^{\mu\nu}=0$, 
we can obtain the following relations:
\begin{equation}
\nabla_\mu \left( \rho \Sigma^{\mu\nu} \right) = 0,\quad \mbox{and} \quad
\Sigma^{\mu\beta} \nabla_\mu 
\left[ \frac{\Sigma_\beta^{\ \nu}}{(-\gamma)^{1/2}} \right] = 0.
\end{equation}

Considering a static and spherically symmetric spacetime of the form \eqref{metricageral} and the Einstein field equations, written as
\begin{equation}
    R_{\mu\nu}-\frac{1}{2}g_{\mu\nu}R=T_{\mu\nu},
\end{equation}
with the above stress-energy tensor, Letelier found an exact BH solution of the form
\begin{equation}
f(r) = 1 - \frac{2M}{r} - \alpha,
\label{eq:letelier_f}
\end{equation}
where $M$ is the ADM mass of the BH and $\alpha$, limited by $0<\alpha<1$, is an integration constant related to the cloud of strings and is related to the bivector $\Sigma^{\mu\nu}$ by $\Sigma^{01}=\alpha/\rho r^2$.

The radius of the event horizon is determined by the condition $f(r_h) = 0$, i.e.,
\begin{equation}
r_h = \frac{2M}{1 - \alpha}.
\label{eq:horizon_radius}
\end{equation}

The stress-energy tensor of the cloud of strings corresponds to a source with vanishing tangential pressures and non-vanishing radial pressure and energy density,
\begin{equation}
T^t{}_t = T^r{}_r = - \frac{\alpha}{ r^2}, \qquad T^\theta{}_\theta = T^\varphi{}_\varphi = 0,
\label{eq:T_components_string}
\end{equation}
which represents a radially oriented distribution of strings. In the limit $\alpha \to 0$, the Schwarzschild solution is recovered.

The regularity of this spacetime can be analyzed through the Kretschmann scalar\footnote{The Kretschmann scalar is defined as
$K=R_{\mu\nu\alpha\beta}R^{\mu\nu\alpha\beta}$, where $R_{\mu\nu\alpha\beta}$ is the Riemann tensor.}, which is written as
\begin{equation}
    K=\frac{4 \left(\alpha^2 r^2+4 \alpha M r+12 M^2\right)}{r^6}.
\end{equation}
It is clear that this spacetime has a curvature singularity at $r=0$.

\subsection{New cloud of strings spacetime: The Letelier-Alencar solution}

One of the key points considered by Letelier to obtain his solution was the fact that $\gamma < 0$, which implies that only the $\Sigma^{01}$ component is nonzero. However, more recently, Alencar and collaborators showed that it is possible to keep the condition $\gamma < 0$ while also allowing a nonvanishing $\Sigma_{23}$ component, thus obtaining a generalized version of the Letelier solution \cite{Alencar:2025zyl}.

In this way, we now consider that the nonvanishing components of the bivector are given by:
\begin{equation}
    \rho \Sigma_{01}=\frac{g_s^2}{r^2}, \quad \rho \Sigma_{23}=b(r) \sin \theta,
\end{equation}
where
\begin{equation}
    b(r)=\frac{g_s^2 l_s^2}{\sqrt{l_s^4+r^4}},
\end{equation}
with $l_s$ being the string length and $g_s^2$ the effective coupling controlling the gravitational contribution of the cloud of strings at tree level\footnote{In fact, the authors initially obtain these components with the constants $\alpha$, as in the Letelier case, and $c_0$. However, through their analysis, they relate the constant $c_0$ to $l_s$ and the constant $\alpha$ to $g_s^2$ \cite{Alencar:2025zyl}.}.

With these results, $\gamma$ is now given by
\begin{equation}
    \gamma=-\frac{g_s^4}{\rho(l_s^4+r^4)}<0.
\end{equation}
With this, we see that the generalization of the cloud of strings preserves the criterion that $\gamma$ is negative. The components of the stress-energy tensor are given by
\begin{equation}
T^t{}_t = T^r{}_r = - \frac{g_s^2}{r^4}\sqrt{l_s^4+r^4}, \quad T^\theta{}_\theta = T^\varphi{}_\varphi = \frac{g_s^2l_s^4}{r^4\sqrt{l_s^4+r^4}}.
\label{eq:T_components_string_gen}
\end{equation}
All these results reduce to Letelier's results in the limit $l_s \to 0$ and $g_s^2 \to \alpha$.

The spacetime found by the authors is described by the line element \eqref{metricageral}, where $f(r)$ is given by
\begin{equation}
f(r) = 1 - \frac{2M}{r} 
+ \frac{g_s^2 l_s^2}{r^2} \, 
{}_2F_{1}\left(-\frac{1}{2}, -\frac{1}{4}, \frac{3}{4}, -\frac{r^4}{l_s^4}\right),\label{geova_solution}
\end{equation}
with ${}_2F_{1}\left(a, b, c, d\right)$ being the hypergeometric function.

By analyzing the function~\eqref{geova_solution}, we can interpret the $l_s$ and $g_s$ corrections as introducing a repulsive contribution to the gravitational potential, which competes with the attractive mass term $M$. The balance between these effects leads to interesting implications for several physical properties of the solution.

\begin{figure*}[htb]
    \centering
    \includegraphics[width=.5\linewidth]{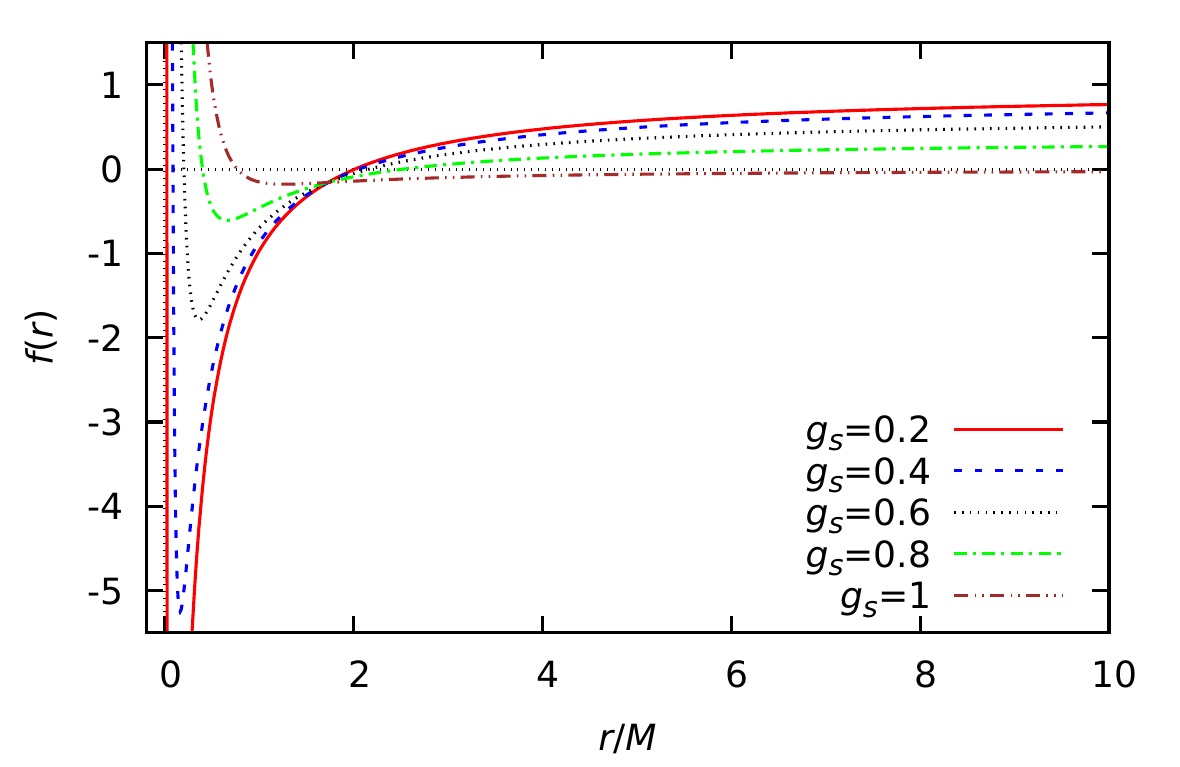}\hspace{-0.1cm}
    \includegraphics[width=.5\linewidth]{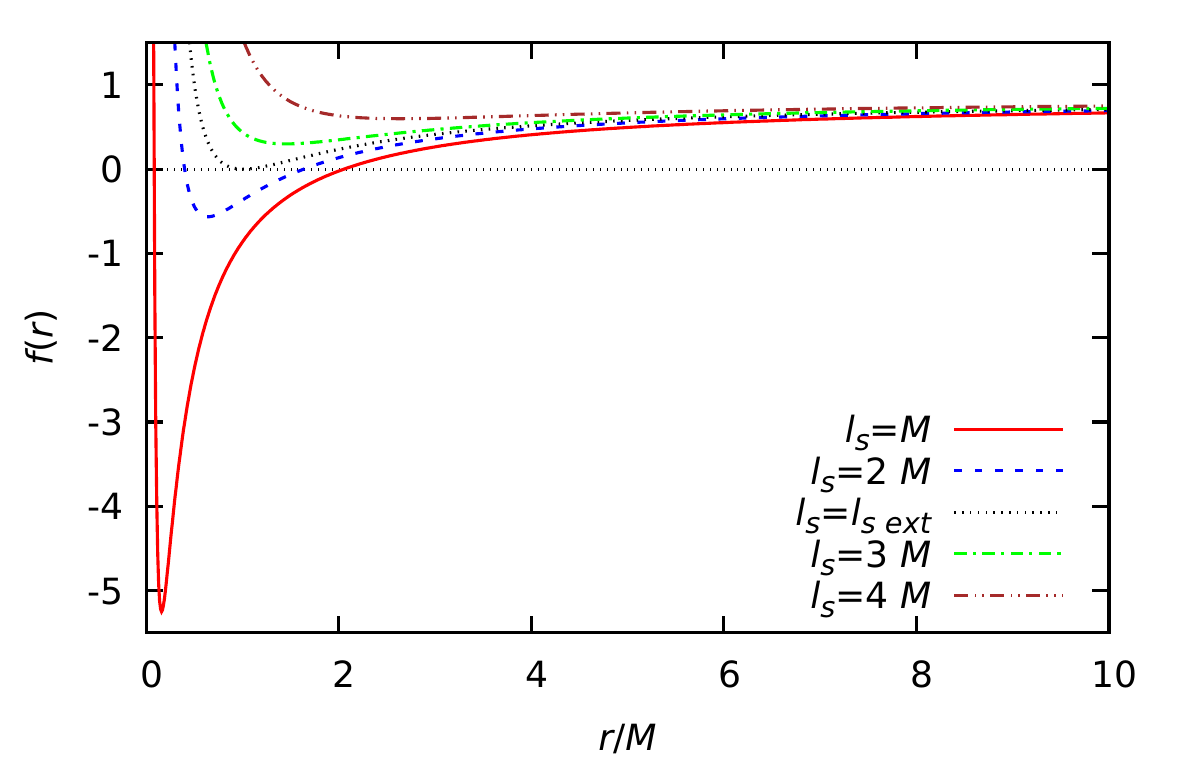}
    \caption{Behavior of the function $f(r)$ as a function of the radial coordinate for different values of $g_s$ with $l_s/M = 1$ (left), and for $g_s = 0.4$ with different values of $l_s$ (right).
}
    \label{fig:f}
\end{figure*}

\begin{figure*}[htb]
    \centering
    \includegraphics[width=.5\linewidth]{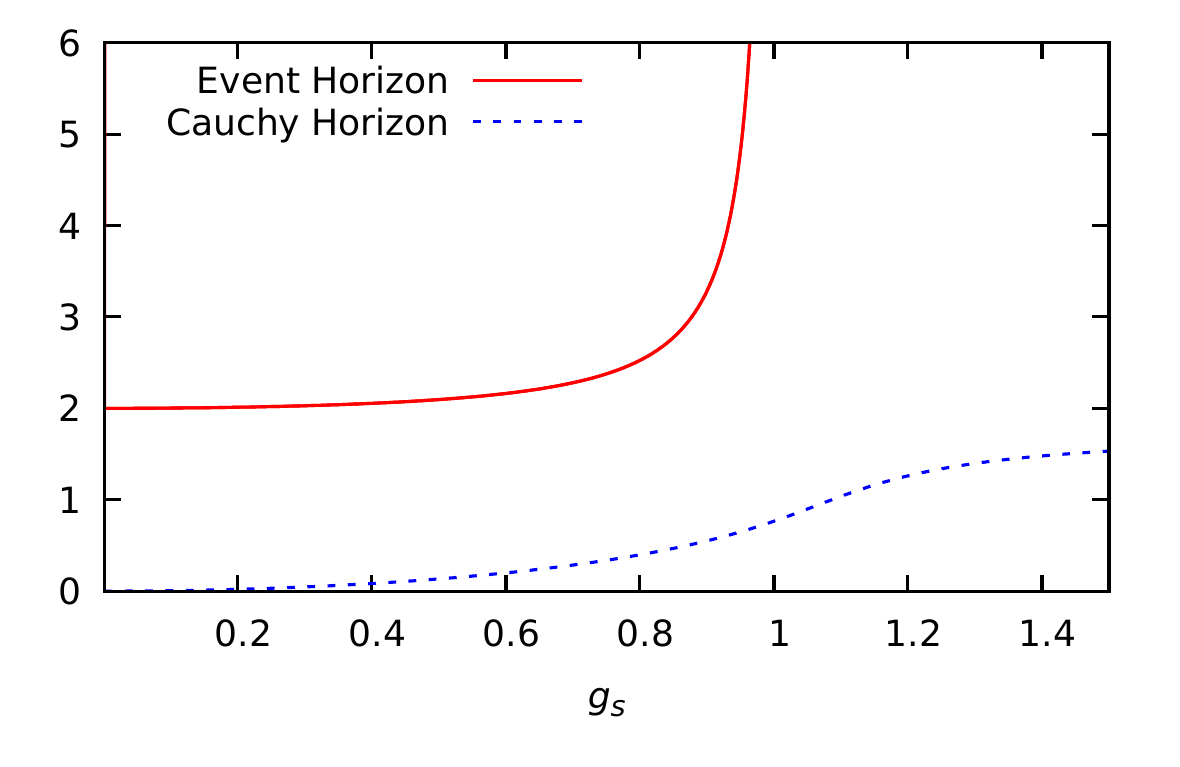}\hspace{-0.1cm}
    \includegraphics[width=.5\linewidth]{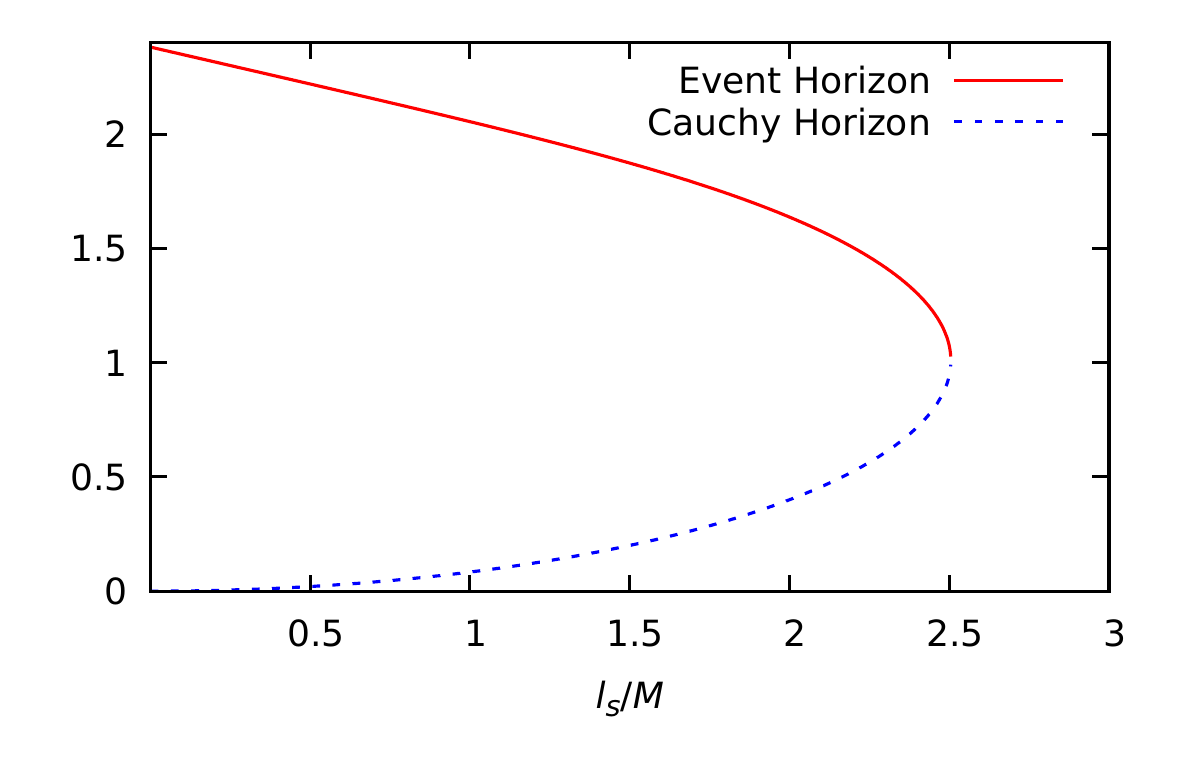}
    \caption{Behavior of the event horizon radius and the Cauchy horizon radius as functions of $g_s$ with $l_s/M = 1$ (left) and as a function of $l_s$ with $g_s=0.4$ (right).
}
    \label{fig:horizon}
\end{figure*}

It is not possible to solve $f(r)=0$ analytically to obtain the horizon radii. However, it is still possible to verify the existence of horizons graphically. In Fig. \ref{fig:f}, we show the behavior of the function $f(r)$ as a function of the radial coordinate for different values of $g_s$. We can see that for $g_s < 1$ there are two horizons, an event horizon and a Cauchy horizon, while for $g_s > 1$ the event horizon ceases to exist and the signature of the metric changes. We also observe that for $g_s < 1$ it is possible to have two horizons, an extremal horizon, or no horizon at all, similar to the Reissner-Nordström case. In \cite{Alencar:2025zyl}, the authors analyze the existence of one or more horizons through the behavior of the derivative of $f(r)$ and conclude that more than one horizon can exist only if $g_s < 1$. Already in their original work, the authors established that the condition $g_s < 1$ is required in order to ensure a positive Hawking temperature. In Fig.~\ref{fig:horizon}, we show the behavior of the event horizon and the Cauchy horizon. The radius of the event horizon tends to diverge as $g_s \to 1$, and for $g_s > 1$ only the Cauchy horizon remains in the solution. For $g_s < 1$, the radius of the event horizon decreases and the radius of the Cauchy horizon increases as $l_s$ grows, until only an extremal horizon remains for $l_s = l_s^{\text{ext}}$. For $l_s > l_s^{\text{ext}}$, there are no horizons.

\begin{figure}[htb]
    \centering
    \includegraphics[width=1\linewidth]{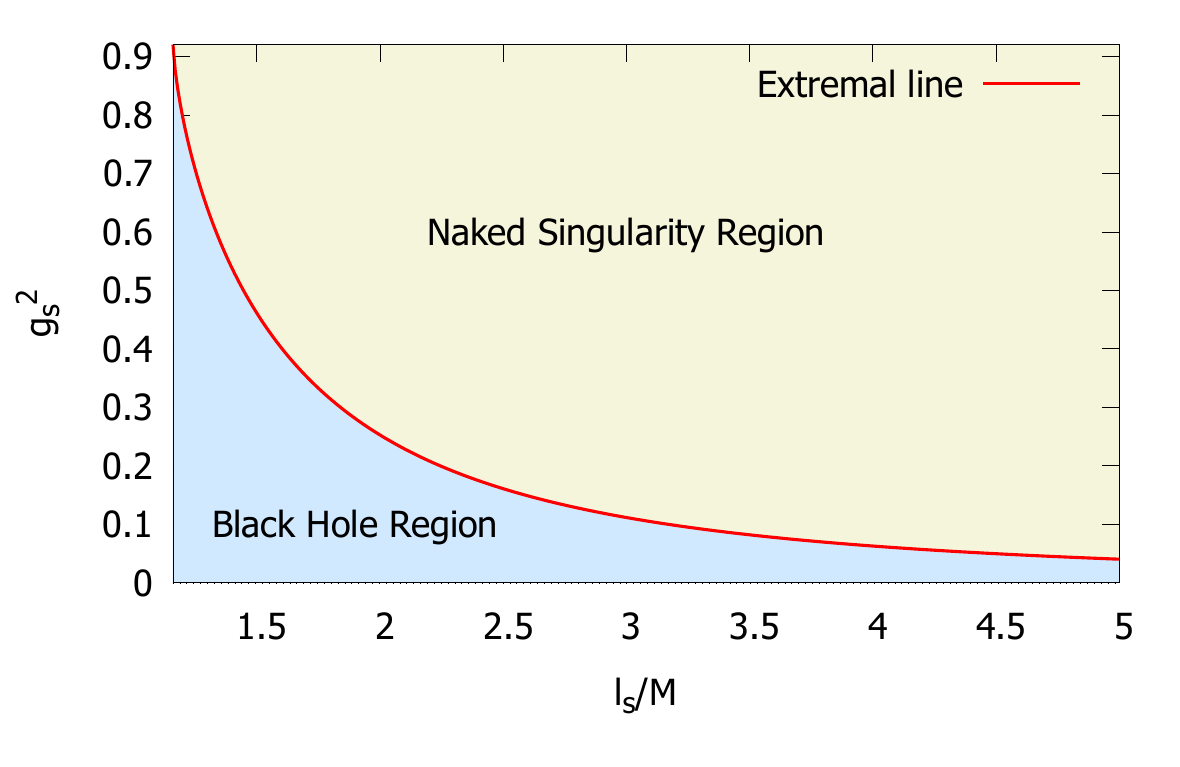}
    \caption{Phase diagram in the $(l_s/M,\, g_s^2)$ plane. The solid red curve represents the extremal limit, where the inner and outer horizons coincide. This curve separates the black hole region (light blue), in which an event horizon cloaks the singularity, from the naked singularity region (beige).
}
    \label{fig:phasediagram}
\end{figure}

To clarify how the parameters of the cloud of strings influence the causal structure of the solution, we present the phase diagram in the $(l_s/M,\, g_s^2)$ plane in Fig.~\ref{fig:phasediagram}. The solid red curve defines the extremal limit , at which the inner and outer horizons degenerate. This boundary separates the physical black hole region (light blue shaded area), where an event horizon shields the singularity, from the naked singularity region (beige area).

From a physical perspective, the diagram reveals that, as the string length $l_s$ increases, the domain of existence of black hole solutions shrinks, requiring a significantly smaller coupling $g_s$ in order to maintain the horizon. This behavior reflects the fact that larger string scales enhance the quantum dispersive effects that counteract gravitational attraction, thereby demanding a weaker coupling to preserve the black hole solution.

To verify the regularity of this spacetime, we analyze the behavior of the Kretschmann scalar around $r \to 0$, which is given by
\begin{equation}
        K \approx \frac{56 g_s ^4 l_s^4}{r^8}-\frac{96 g_s^2  l_s^2 M}{r^7}+\frac{48 M^2}{r^6}+O\left(\frac{1}{r^5}\right), \quad \mbox{if} \quad r\rightarrow 0.
\end{equation}
Thus, we see that the generalization of the Letelier solution exhibits even more divergent terms as $r \to 0$.

In the next section, we will study the geodesics in these spacetimes and then use them to analyze the tidal forces.

\section{Geodesics}\label{SEC:geodesics}
Geodesics are fundamental for studying tidal forces once they describe the trajectories of massive particles and photons in different gravitational regimes. By analyzing the deviation between nearby geodesics, one can determine how spatial separations evolve along radial or orbital paths, providing a direct description of tidal forces in the spacetime.

To obtain geodesics in the spacetime described by the line element \eqref{metricageral}, we can use the Lagrangian associated with the metric, which is written as
\begin{equation}
    \mathcal{L}=\frac{1}{2}\left(-f(r)\dot{t}^2 + \frac{\dot{r}^2}{f(r)} + r^2\dot{\theta}^2+r^2\sin^2\theta \dot{\varphi}^2\right),
\end{equation}
where the dot represents the derivative with respect to an affine parameter $\tau$.

Since the Lagrangian does not explicitly depend on $t$ and $\varphi$, we have the following conserved quantities
\begin{equation}\label{conserved}
    E=-\frac{\partial \mathcal{L}}{\partial \dot{t}}=f \dot{t}, \quad \mbox{and} \quad L=\frac{\partial \mathcal{L}}{\partial\dot{\varphi}}=r^2\sin^2\theta \dot{\varphi},
\end{equation}
where $E$ and $L$ are interpreted as the energy and angular momentum of the particle, respectively.

In addition to the relations above, to describe the geodesics we also have the following relation
\begin{equation}\label{geo}
    -f(r)\dot{t}^2 + \frac{\dot{r}^2}{f(r)} + r^2\dot{\theta}^2+r^2\sin^2\theta \dot{\varphi}^2=-\delta,
\end{equation}
where $\delta = 0$ for massless particles and $\delta = 1$ for massive particles.

The equation above can be simplified in two different ways, depending on the type of motion we want to analyze.  
The first case corresponds to the motion of particles falling radially into a BH. This implies $\dot{\theta} = \dot{\varphi} = 0$, so that the equation above, combined with \eqref{conserved}, is simplified as
\begin{equation}\label{georad}
    E^2  =\dot{r}^2 +f(r)\delta.
\end{equation}
The other way to simplify Eq.~\eqref{geo} is by considering particles moving in the equatorial plane, i.e., $\theta = \pi/2$.  
In this case, together with Eq.~\eqref{conserved}, Eq.~\eqref{geo} becomes
\begin{equation}E^2=\dot{r}^2+f(r)\left(\frac{L^2}{r^2}+\delta\right).
\end{equation}
Alternatively, we can write this equation as
\begin{equation}\label{geocirc}
E^2=\dot{r}^2+V_{eff},
\end{equation}
with effective potential $V_{eff}$ being
\begin{equation}\label{pot_gen}
   V_{eff}= f(r)\left(\frac{L^2}{r^2}+\delta\right).
\end{equation}

We now consider separately the massive and massless cases, as well as the different types of motion.

\subsection{Photons}
For photons, we will focus on the formation of light rings, which are characterized by circular photon orbits. Accordingly, we will ignore the purely radial motion and consider only motion in the equatorial plane.

For massless particles, we have $\delta = 0$ in Eq.~\eqref{pot_gen}, and therefore the effective potential for photons is given by
\begin{equation}\label{potmassless}
   V_{ph}= \frac{f(r)L^2}{r^2}.
\end{equation}
This is the potential experienced by massless particles moving in this spacetime. The potential vanishes at the points where $f(r) = 0$, i.e., at the horizons. 
Circular orbits are characterized by $\dot{r} = \ddot{r} = 0$. 
Therefore, the radius of the photon sphere can be determined from the conditions
\begin{equation}\label{LR_conditions}
    \left.V_{ph}\right|_{r=r_{LR}}=E^2, \quad \mbox{and} \quad \left.\frac{dV_{ph}}{dr}\right|_{r=r_{LR}}=0.
\end{equation}
We define $r_{\rm LR}$, the radius of the light ring, as the point where $V_{\rm ph}$ reaches its maximum, so that the photon has an unstable circular orbit.

\begin{figure*}[htb]
    \centering
    \includegraphics[width=.5\linewidth]{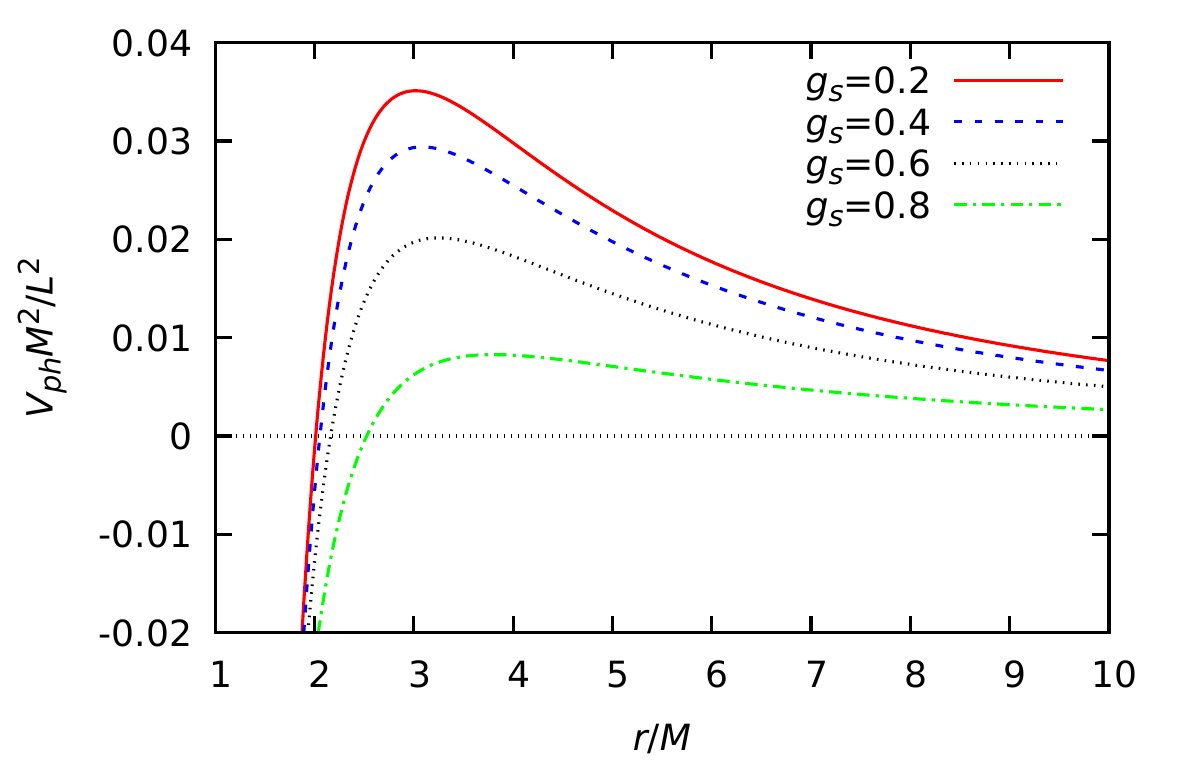}\hspace{-0.1cm}
    \includegraphics[width=.5\linewidth]{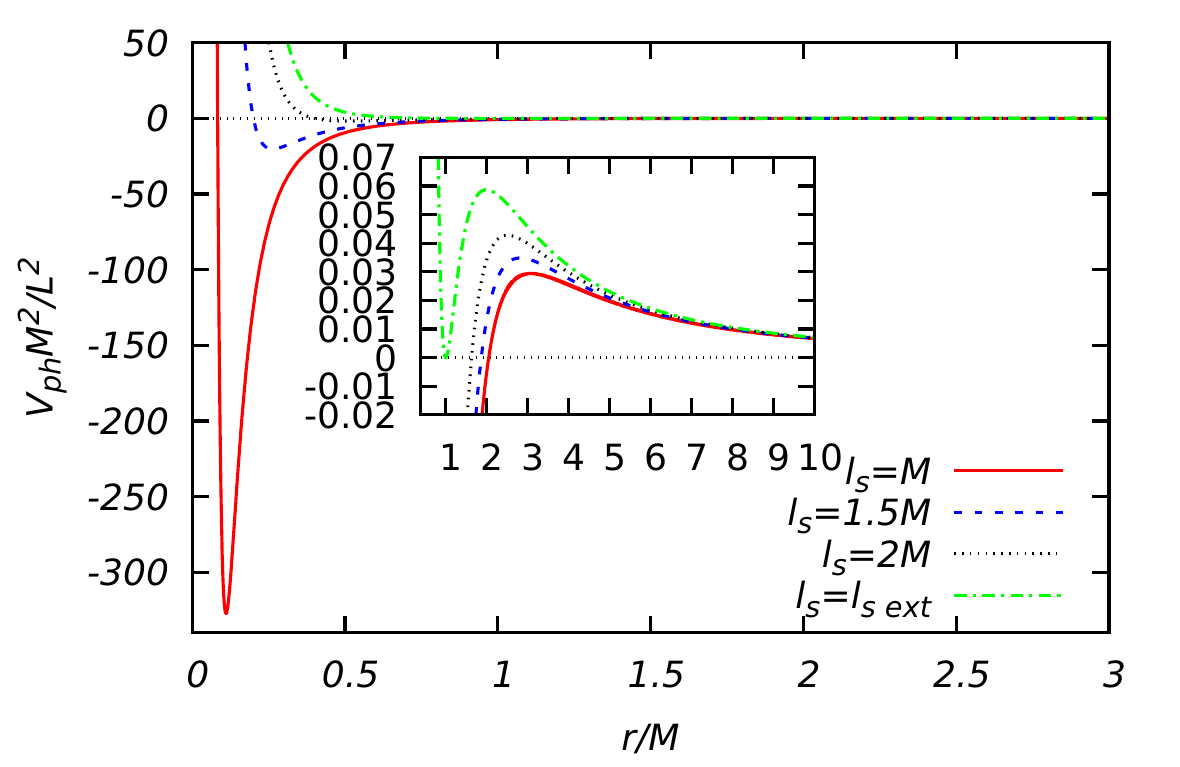}
    \caption{Behavior of the effective potential for photons as a function of the radial coordinate for $l_s/M = 1$ and different values of $g_s$ (left) and for $g_s = 0.4$ with different values of $l_s$ (right).
}
    \label{fig:Vmassless}
\end{figure*}

In Fig.~\ref{fig:Vmassless}, we present the behavior of the effective potential for photons. 
Its magnitude increases as $l_s$ increases and decreases as $g_s$ increases. 
The location of the maximum does not appear to be strongly affected by variations in $g_s$, only for small values of $g_s$, while it is more sensitive to changes in $l_s$. The potential maximum corresponds to unstable circular photon orbits. A minimum is also present, but it occurs between the horizons; 
in that region, the metric signature changes and the coordinate $r$ becomes timelike. 
Therefore, the most relevant feature is the potential maximum.

\begin{figure*}[htb]
    \centering
    \includegraphics[width=.5\linewidth]{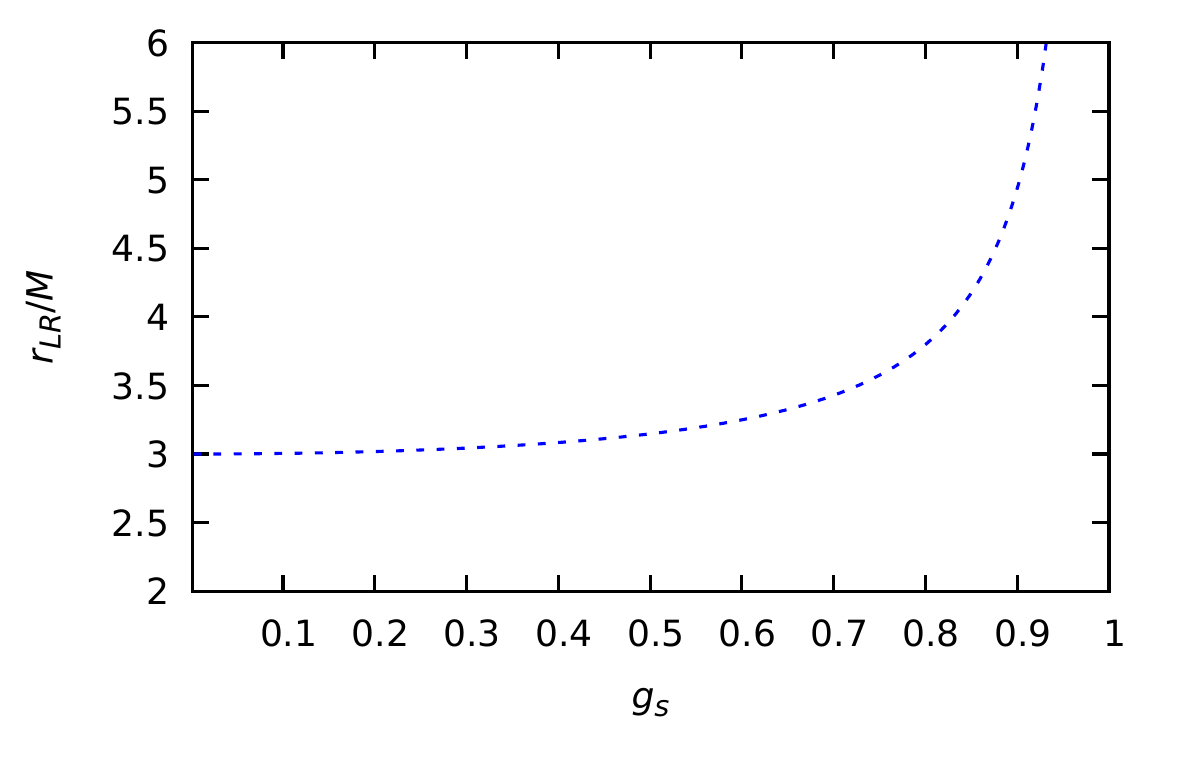}\hspace{-0.1cm}
    \includegraphics[width=.5\linewidth]{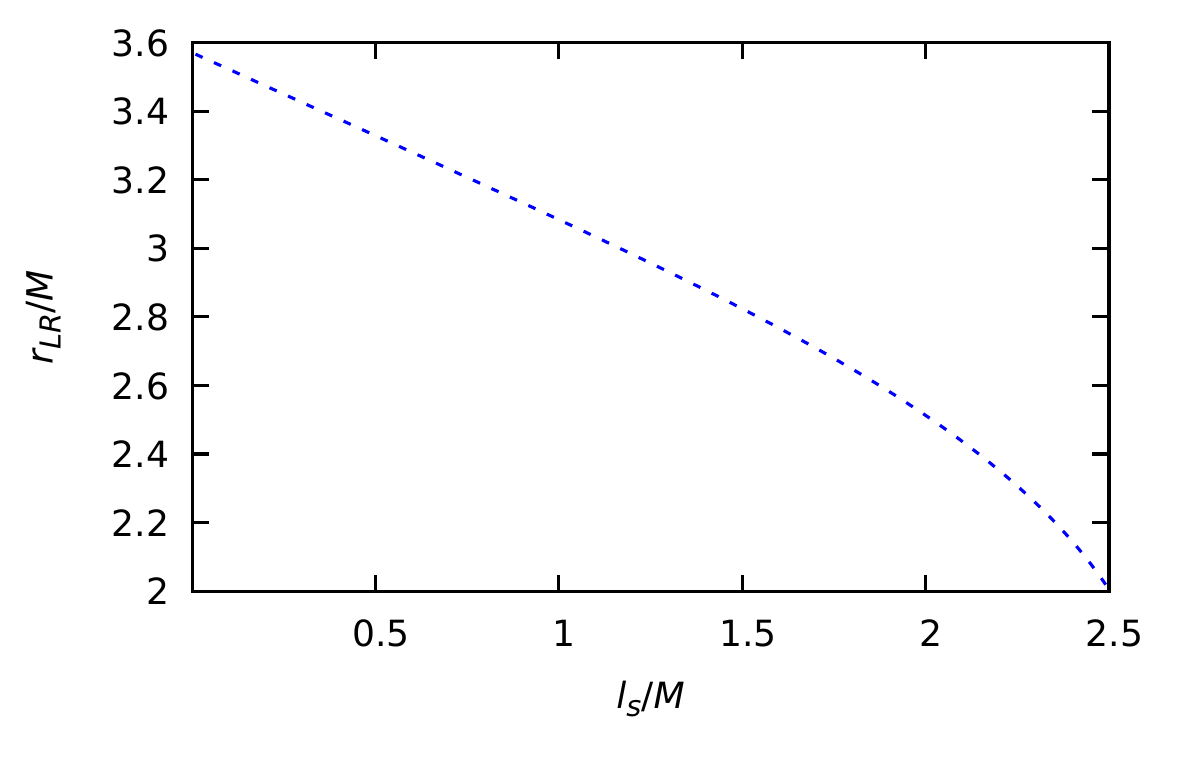}
    \caption{Behavior of the unstable photon orbit as a function of $g_s$ with $l_s/M=1$ (left panel) and as a function of $l_s$ for $g_s=0.4$ (right panel).
}
    \label{fig:LR}
\end{figure*}

From conditions \eqref{LR_conditions}, we obtain the following expression:
\begin{equation}\label{LR_eq}
    r_{LR}f'(r_{LR})-2f(r_{LR})=0.
\end{equation}
Solving this expression yields the photon orbit radius. However, for our model, the equation cannot be solved analytically; therefore, we will solve it numerically and analyze the results through plots. In Fig.~\ref{fig:LR}, we show the behavior of the unstable photon orbit radius. As noted above, for small values of $g_s$ there are no significant changes; however, as $g_s$ approaches $1$, the radius increases abruptly. When varying $l_s$, the maximum orbit radius occurs for $l_s \to 0$ and decreases as $l_s$ grows.

\subsection{Massive particles in equatorial plane}

\begin{figure}[htb]
    \centering
    \includegraphics[width=1\linewidth]{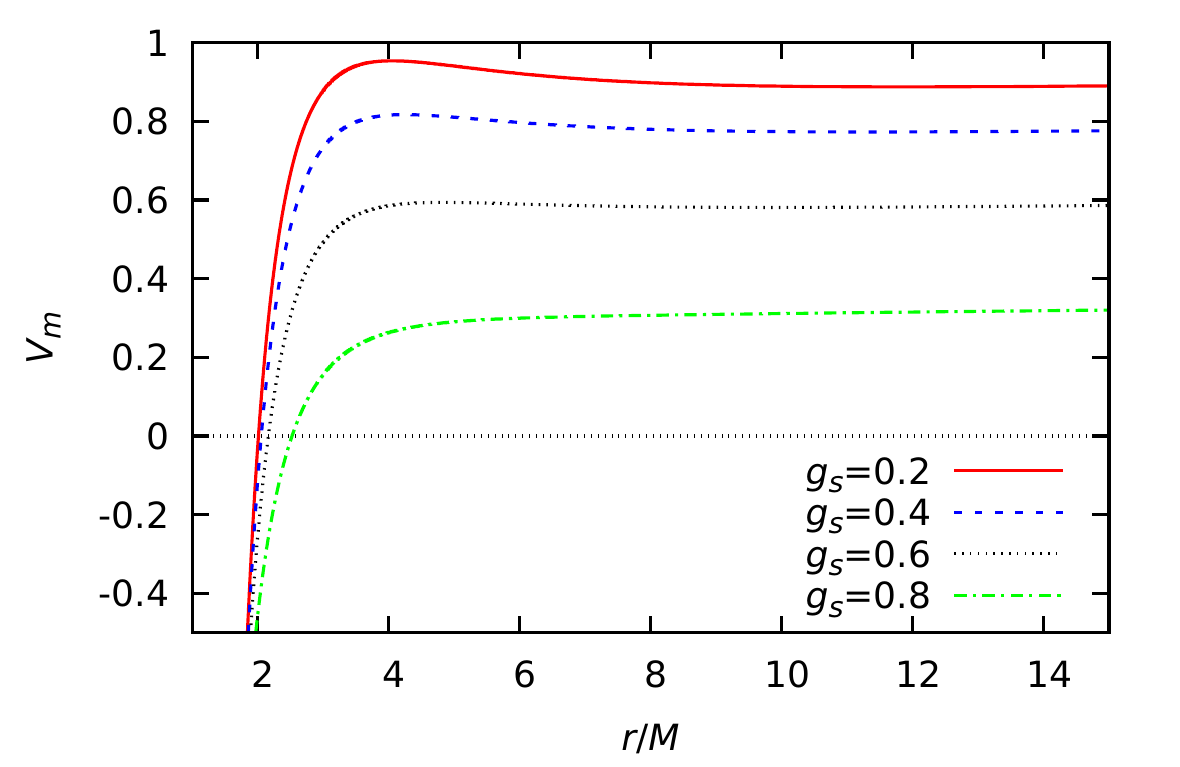}
    \includegraphics[width=1\linewidth]{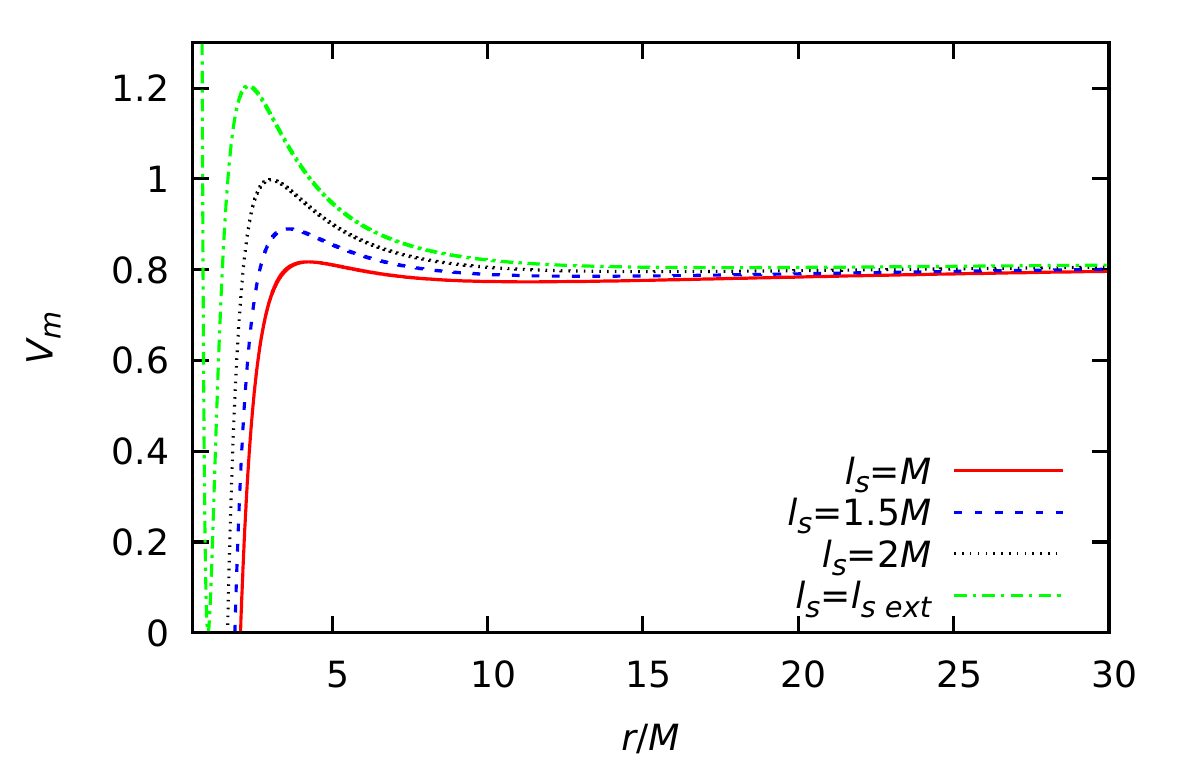}
    \includegraphics[width=1\linewidth]{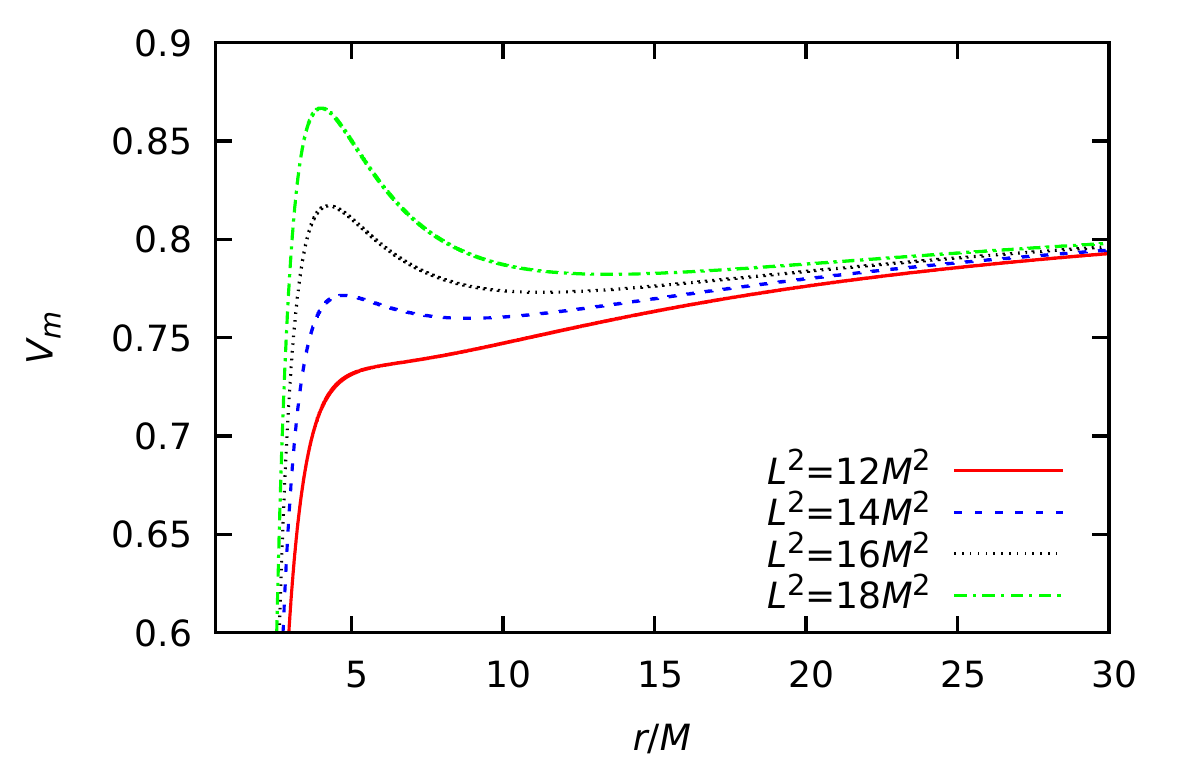}
    \caption{Effective potential for massive particles under different parameter choices. Top panel: $L/M=4$, $l_s/M=1$, varying $g_s$. Middle panel: $L/M=4$, $g_s=0.4$, varying $l_s$. Bottom panel: $g_s=0.4$, $l_s/M=1$, varying $L$.}
    \label{fig:pot_m}
\end{figure}

For massive particles in the equatorial plane, the effective potential can be written as
\begin{equation}
    V_{m}=f(r)\left(\frac{L^2}{r^2}+1\right).
\end{equation}
It is important to note that, as in the massless case, the effective potential for massive particles also vanishes at the horizons. In the limit $r \to \infty$, the effective potential for massive particles approaches $\lim_{r\to\infty}V_m = 1 - g_s^2$, whereas it is zero for massless particles. In Fig.~\ref{fig:pot_m}, we show the behavior of the effective potential for different parameter choices. Outside the event horizon there are both maxima (unstable orbits) and minima (stable orbits). In the massive case, $L$ plays a crucial role, since depending on its value, circular orbits may not exist.

For circular orbits, we impose $\dot{r}=\ddot{r}=0$. From these conditions, we can determine the energy and angular momentum associated with circular orbits, given by
\begin{equation}\label{E_L_massive}
    E_c^2=\frac{2f(r_c)^2}{2f(r_c)-f'(r_c)r_c}, \quad L_c^2=\frac{r_c^3 f'(r_c)}{2f(r_c)-f'(r_c)r_c},
\end{equation}
where $r_c$ is the radius of the circular orbit. By comparing Eq.  \eqref{E_L_massive} with Eq. \eqref{LR_eq}, we see that the energy and the angular momentum diverge precisely as the radius of the circular orbit for massive particles approaches the radius of the circular orbit for massless particles.

To find the radius of circular orbits, we solve $V_m'(r_c)=0$, and their stability is assessed by the sign of $V_m''(r_c)$. From Fig.~\ref{fig:pot_m}, we see that the stable orbit ($V_m''(r_c)>0$) has a larger radius than the unstable orbit ($V_m''(r_c)<0$). For the generalized solution, it is not possible to obtain the radii of these orbits analytically. However, they can be computed numerically, and we can analyze their behavior as the solution parameters vary.

\begin{figure}[htb]
    \centering
    \includegraphics[width=1\linewidth]{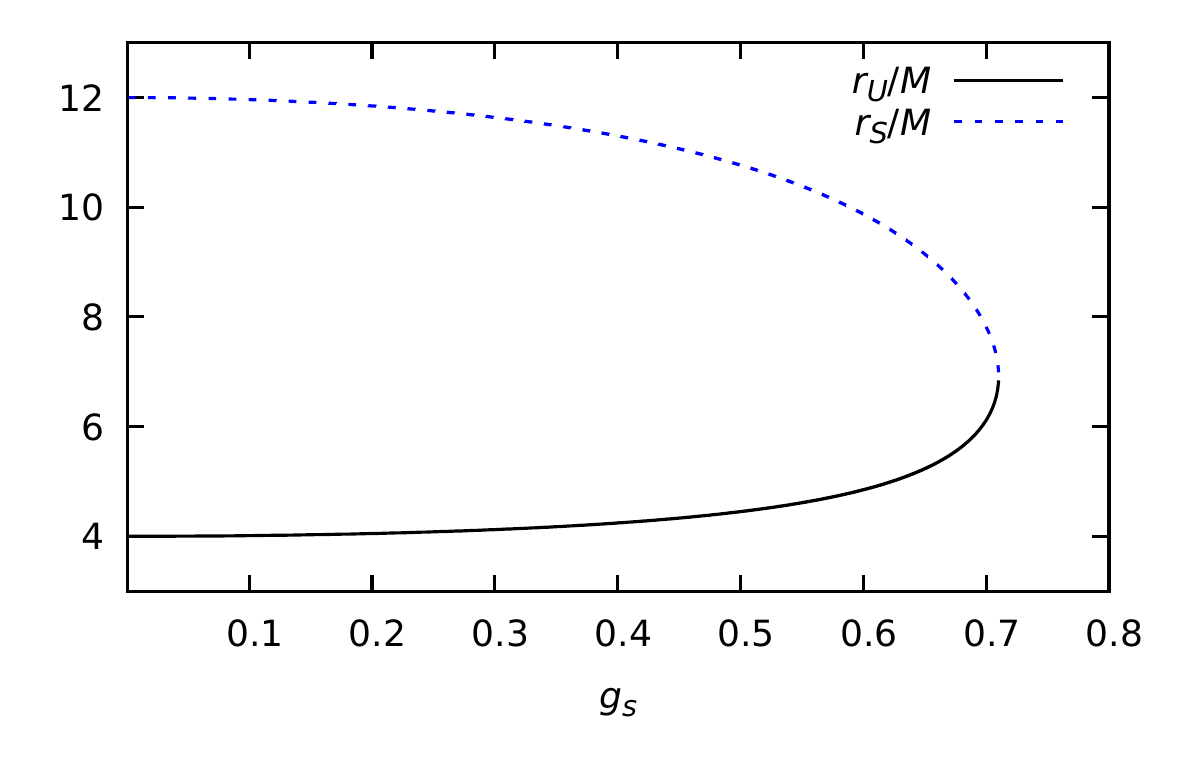}
    \includegraphics[width=1\linewidth]{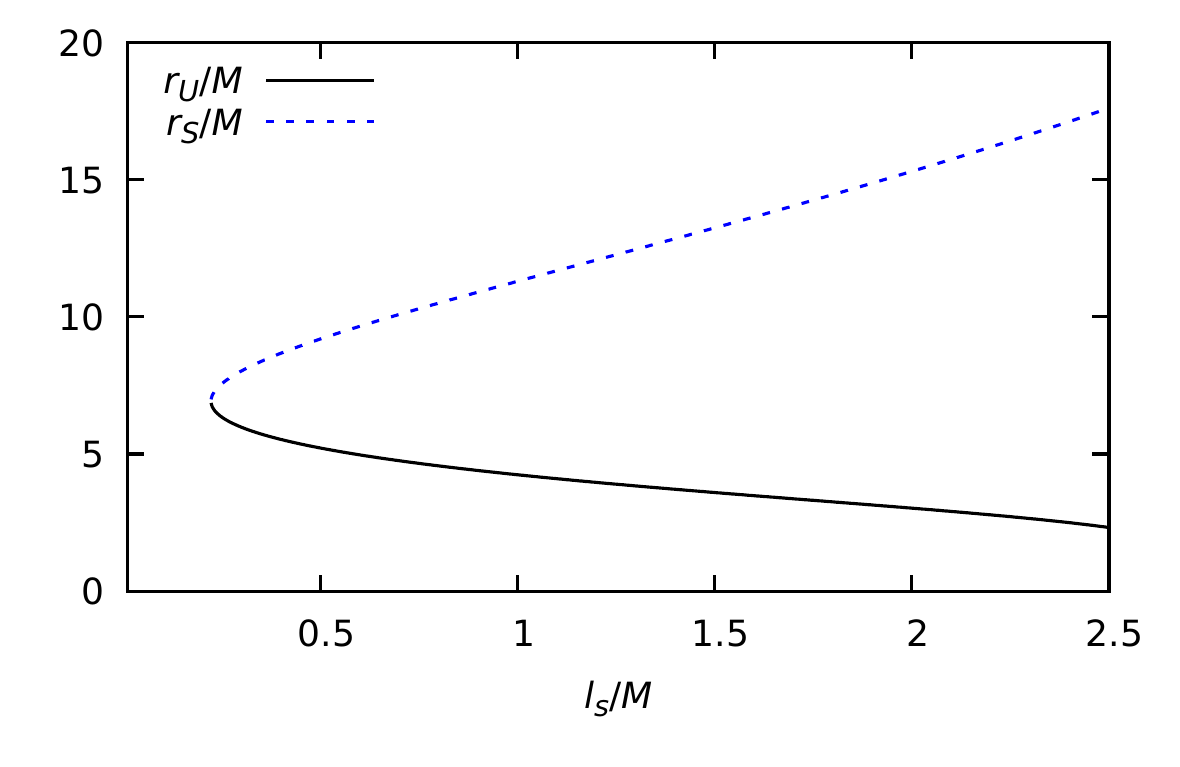}
    \includegraphics[width=1\linewidth]{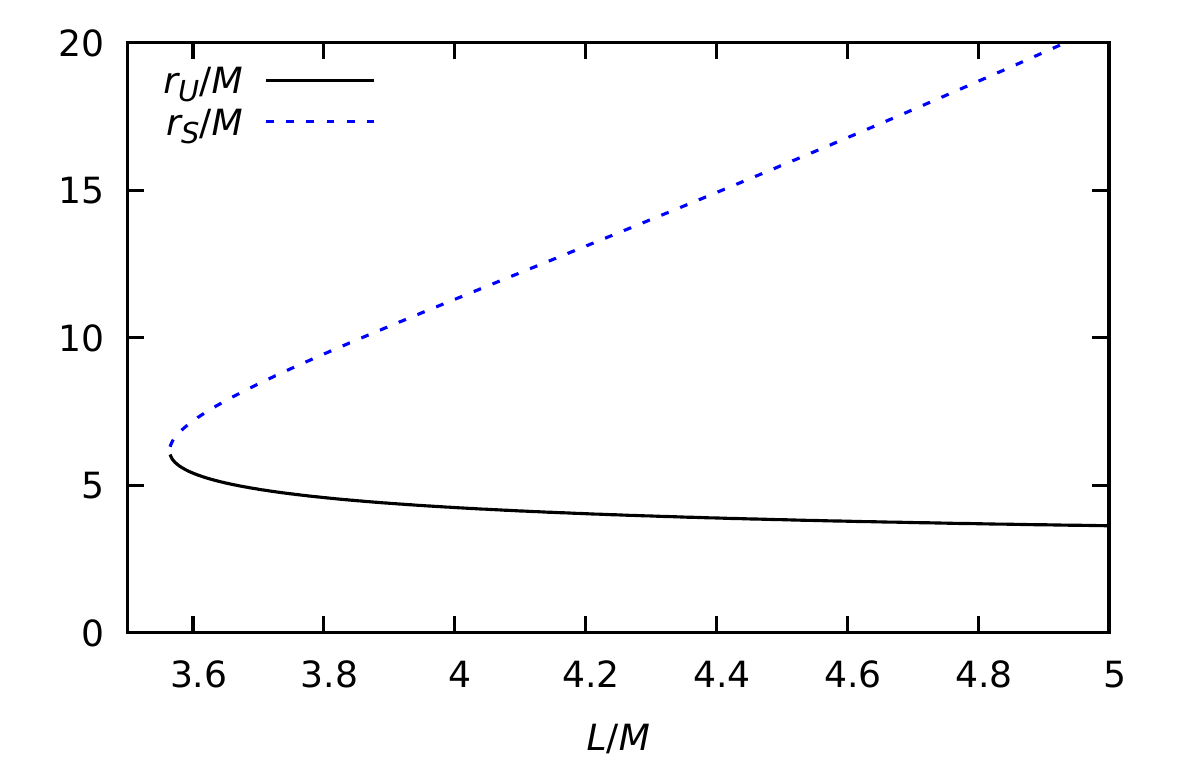}
    \caption{Behavior of the stable orbit radius $r_S$ and the unstable orbit radius $r_U$ for massive particles, considering different parameter values. In the top panel, we fix $L=4M$ and $l_s=M$ and vary $g_s$. In the middle panel, we fix $g_s=0.4$ and $L=4M$ and vary $l_s$. In the bottom panel, we fix $g_s=0.4$ and $l_s=M$ and vary $L$.}
    \label{fig:raio_C_M}
\end{figure}

Figure \ref{fig:raio_C_M} shows the behavior of the stable and unstable orbit radii for massive particles, considering different parameter values. We observe that, as $g_s$ increases, the radii tend to the same value. For $g_s \to 0$, we recover the Schwarzschild results. When analyzing the dependence on $l_s$, a minimum value of $l_s$ is required for these orbits to exist. This occurs because, in the Letelier limit ($l_s \to 0$), the orbits exist only if $g_s < \sqrt{1 -2\sqrt{3}M/L}$. Therefore, for the $g_s$ value chosen in the plot, $l_s$ cannot take very small values. As a function of $L$, there exists a minimum value for which the circular orbit radii exist. It is well known that, as $L$ decreases, a critical value is reached at which the maximum and minimum of the potential coalesce, producing an inflection point. At this point, the radius equals that of the innermost stable circular orbit (ISCO). 

From conditions $V'_m=0$ and $V''_m=0$, we obtain the following expression:
\begin{equation}
2r_{I}f(r_{I})f''(r_{I})-4rf'(r_{I})^2+6f(r_{I})f'(r_{I})=0,
\label{eq:cond-isco-en}
\end{equation}
where $r_I$ is the radius of the ISCO. If we solve \eqref{eq:cond-isco-en} to the Schwarzschild case, we find $r_I=6M$. To the Letelier solution, we get $r_I=6 M/(1-g_s^2)$.

\begin{figure*}[htb]
    \centering
    \includegraphics[width=.5\linewidth]{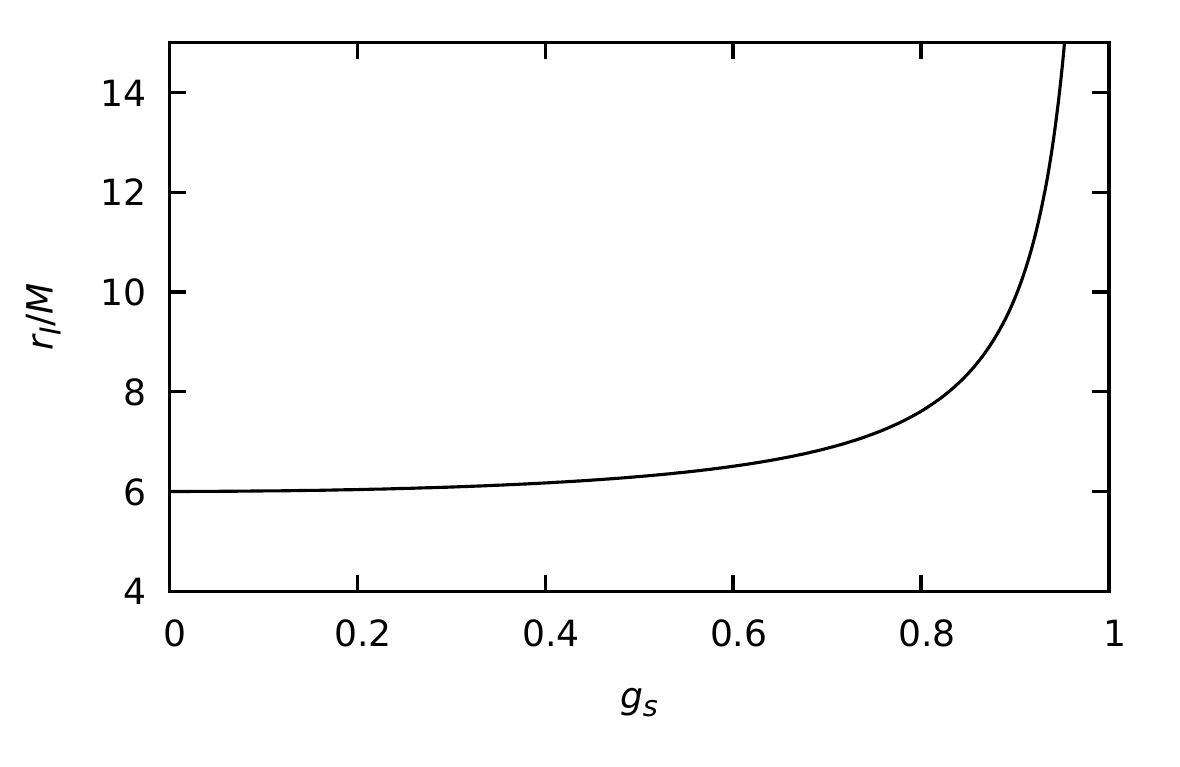}\hspace{-0.1cm}
    \includegraphics[width=.5\linewidth]{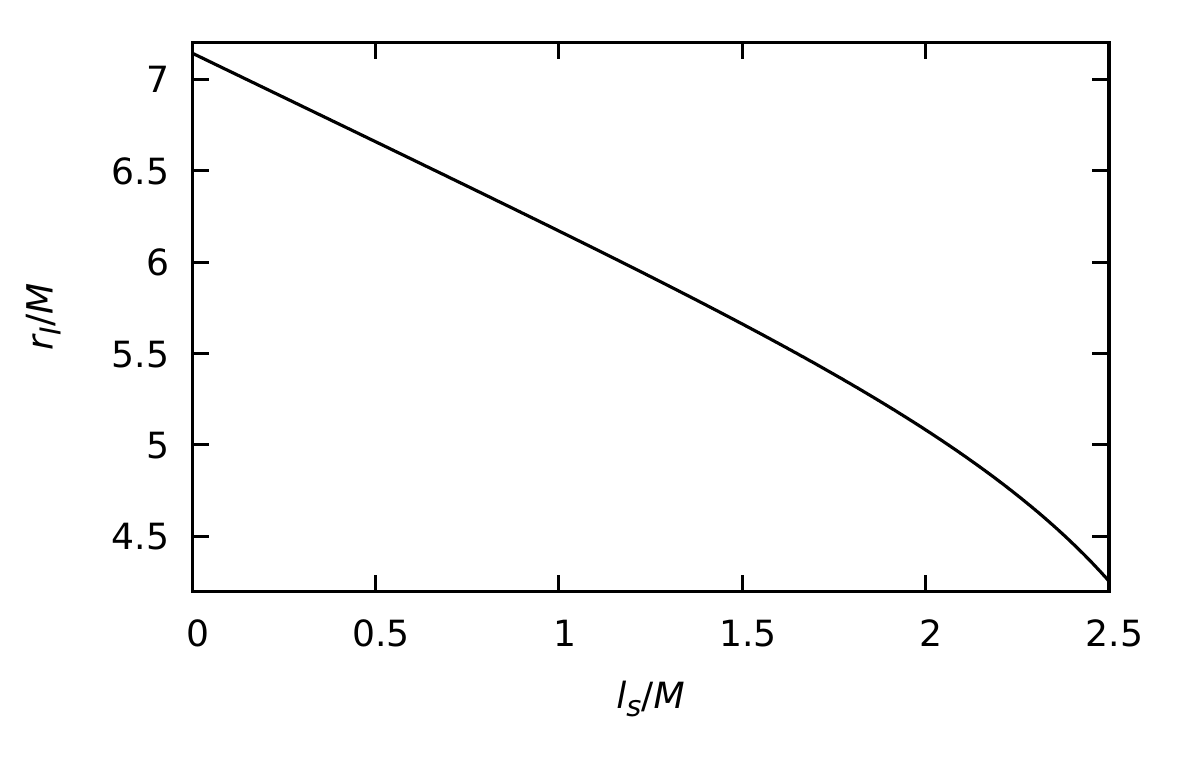}
    \caption{Behavior of the ISCO radius for $l_s = M$ varying $g_s$ (left panel) and for $g_s = 0.4$ varying $l_s$ (right panel).}
    \label{fig:isco}
\end{figure*}

In Fig.~\ref{fig:isco}, we observe how the ISCO radius behaves as the parameters vary. For small values of $g_s$, there is little change in the radius, but as $g_s \to 1$, the radius increases abruptly. The dependence on $l_s$ is smoother and decreases as $l_s$ increases.

\subsection{Massive particles in radial motion}
For radial motion, we consider only massive particles ($\delta=1$), so that the relation \eqref{georad} is simply
\begin{equation}
    E^2=  \dot{r}^2+f(r),\quad
\mbox{or}\quad
    \label{rgeodesica}  \frac{dr}{d\tau} = \pm \sqrt{E^2 - f(r)}.
\end{equation}

Since we will assume that the body is moving radially toward the BH, we will only consider the negative sign of the equation above. Moreover, if the particle is released from rest at $r = b$, its energy per unit mass will then be given by $E = \sqrt{f(r = b)}$.  

If we wish to determine the acceleration ``felt'' by massive particles in this spacetime, we can define the ``Newtonian radial acceleration'' as $A^N=\ddot{r}$, which is given by
\begin{equation}
\begin{split}
        A^N=-\frac{f'(r)}{2}=-\frac{M}{r^2}+\frac{g_s^2 l_s^2
   \sqrt{\frac{r^4}{l_s^4}+1}}{2 r^3}\\+\frac{g_s^2 l_s^2}{2 r^3} \, _2F_1\left(-\frac{1}{2},-\frac{1}{4};\frac{3}{4};-\frac{r^4}{l_s^4}\right).
   \end{split}
\end{equation}
The first term of this expression is the analog of Newtonian gravity, while the remaining terms are relativistic corrections induced by the presence of the cloud of strings.

\section{Geodesic deviation and tidal forces}\label{SEC:TF}
In this section, we will examine in detail the effects of the  cloud of strings on the tidal forces experienced by a body in the vicinity of a BH. To this end, our discussion will first be based on the general equations that describe tidal forces, considering both the case of a freely falling body and that of a body in circular motion around the BH.

However, before proceeding, it is useful to recall how the Newtonian deviation arises. 
That is, given a reference curve $Y^i(t)$ that satisfies the Newtonian equation of motion 
\begin{equation}
    \ddot{Y}^i = -\partial^iU(Y),
\end{equation}
where $U(Y)$ is the Newtonian potential, $U(Y) = -\frac{M}{r}$, and $Y^i =(x , y, z)$.
We consider a second curve $X^i(t)$ constructed as a small deviation from the reference one,
\begin{equation}
    X^i(t) = Y^i(t) + \xi^i(t),
\end{equation}
where $\xi^i(t)$ represents the deviation curve between the two trajectories. To first order in $\xi^i$, the perturbed trajectory $X^i(t)$ must also satisfy the Newtonian equation of motion. 
Expanding the potential around the reference curve $Y(t)$, we obtain
\begin{equation}
    \ddot{X}^i = \partial^i U(X) \simeq \partial^i U(Y) + \partial^i \partial_j U(Y)\, \xi^j.
\end{equation}

Since $Y^i(t)$ already satisfies $\ddot{Y}^i = \partial^i U(Y)$, subtracting the two equations yields
\begin{equation}
    \ddot{\xi}^i(t) = \partial^i \partial_j U(Y)\, \xi^j(t) \equiv  K^i_j\xi^j(t),
\end{equation}
which is the Newtonian deviation equation. If we consider the deviation in the vicinity of a spherically symmetric source of mass $M$, 
it is necessary to perform a coordinate transformation from Cartesian coordinates $x^i$ 
to spherical coordinates $\bar{x}^i$. Because of the spherical symmetry of the system, we can, without loss of generality, 
restrict the reference trajectory to lie in the equatorial plane, defined by 
$\theta = \pi/2$. 
Applying the coordinate transformation $x^i \rightarrow \bar{x}^{i}$ 
from Cartesian to spherical coordinates, 
the Newtonian deviation equation takes the form \cite{Philipp:2016gyq}
\begin{eqnarray}
    &&\ddot{\bar{\xi}}^{i}\, \bar{\partial}_{j} \bar{x}^{i}
    + 2\, \dot{\bar{\xi}}^{i}\, \bar{\partial}_{j} \dot{\bar{x}}^{i}
    + \bar{\xi}^{i}\, \bar{\partial}_{j} \ddot{\bar{x}}^{i}
    \nonumber\\
    &=&
    \Big[
    \partial_r U(r)\, \bar{\partial}^{i} \bar{\partial}_{j} r
    + 
    \partial_r^2 U(r)\,
    (\bar{\partial}^{i} r)(\bar{\partial}_{j} r)
    \Big]
    \bar{\xi}^{k}\, \bar{\partial}_{k} \bar{x}^{j},
\end{eqnarray}
where $\bar{\xi}^{i} = (\xi^r, \xi^{\theta}, \xi^{\varphi})$ 
are the components of the deviation in the new coordinate system, 
and $\bar{\partial}_{i} = (\partial_r, \partial_{\theta}, \partial_{\varphi})$. 
These correspond to three coupled equations for the three unknown components of the deviation. 
All angular terms can be eliminated through appropriate combinations of these equations, 
and after a straightforward but somewhat lengthy calculation, 
one obtains the system of differential equations governing the Newtonian deviations in spherical symmetry. This system of equations, for circular orbits in the equatorial plane ($\dot{r} = \ddot{r} = 0, \, \theta=\pi/2$), is given by
\begin{eqnarray}
    \label{eq:n1}\ddot{\xi}^{\theta} &=& -\,\omega_{K}^{2}\, \xi^{\theta}, \label{eq:xi_theta}\\
    \label{eq:n2}\ddot{\xi}^{r} &=& 2\,r_c\,\omega_{K}\,\dot{\xi}^{\varphi} + 3\,\omega_{K}^{2}\,\xi^{r}, \label{eq:xi_r}\\
   \label{eq:n3} \ddot{\xi}^{\varphi} &=& -\frac{\,2\,\omega_{K}}{r_c}\,\dot{\xi}^{r}. \label{eq:xi_phi}
\end{eqnarray}
The system above corresponds to the Newtonian deviation equations for nearby particles in circular motion around a spherically symmetric source 
with potential $U(Y) = -M/r$. 
In this case, the orbital angular velocity is given by the Keplerian frequency
\begin{equation}
    \omega_{K}^{2} = \frac{M}{r_c^{3}},
\end{equation}
which follows directly from the balance between the gravitational and centrifugal forces 
in a circular orbit.
The equation for the $\theta$-component
has the form of a harmonic oscillator. 
This means that small deviations out of the equatorial plane lead to bounded oscillations with frequency $\omega_{K}$, indicating that the motion is stable under infinitesimal perturbations in the polar direction. 
In other words, if a particle is slightly displaced from the plane of motion, it will oscillate around it without drifting away. Next, we turn our attention to the relativistic case.

\subsection{Relativistic tidal force effects in radial geodesics}
It is well established that, for two infinitesimally nearby geodesics separated by a deviation four-vector $\xi^\gamma$, the relative acceleration between them is described by the geodesic deviation equation \cite{dinverno1992,hobson2006},
\begin{equation}\label{deviation}
    \frac{D^2\xi^\mu}{D\tau^2} = K^\mu_{\gamma}\; \xi^\gamma,
\end{equation}
where $\xi^\mu$ is the infinitesimal displacement vector between nearby geodesics and $K^\mu_{\gamma}$ is the tidal tensor given in terms of the Riemann tensor by
\begin{equation}
    K^\mu_{\gamma} = R^\mu_{\;\;\,\alpha\beta\gamma}u^{\alpha}u^\beta,
\end{equation}
where $u^\mu$ is the timelike unit vector tangent to the geodesic, which satisfies $u^\alpha u^\beta g_{\alpha\beta} = - 1$.

To analyze the effects of tidal forces, we use the tetrad formalism 
\cite{crispim2025, Crispino:2016pnv, Hong:2020tidal}. 
Tetrads are geometric objects that form a set of locally defined orthonormal bases, that is, a set of four linearly independent vectors $\{\vec{e}_{\hat{a}}\}$ defined at each spacetime point. Along the world-line of an observer, these tetrads constitute a local inertial reference frame in which the laws of special relativity hold locally. Since an infinite number of orthonormal frames can be defined at a single point through Lorentz transformations, 
they do not by themselves provide complete information about the spacetime connection. To relate the orthonormal basis to the coordinate basis $\{\vec{e}_{\mu}\}$, 
we introduce the transformation
\begin{equation*}
\vec{e}_{\mu} = \hat{e}^{\;\;\hat{a}}_{\mu}\, \vec{e}_{\hat{a}},
\end{equation*}
where $\vec{e}_{\mu}$ denotes the coordinate basis vectors, 
$\vec{e}_{\hat{a}}$ represents the orthonormal tetrad basis, 
and $\hat{e}^{\;\;\hat{a}}_{ \mu}$ are the tetrad components. In this context, the metric tensor can be expressed in terms of the tetrad vectors as
\begin{equation}
    g_{\mu\nu} = \eta_{\hat{a}\hat{b}}\hat{e}_{\mu}^{\;\; \hat{a}}\hat{e}_{\nu}^{\;\; \hat{b}},
\end{equation}
where the hatted indices are the tetrad basis indices, and $\eta_{\hat{a}\hat{b}} = \text{diag}(-1,\; 1,\; 1,\; 1)$ is the Minkowski metric in Cartesian coordinates.

The tetrad basis $\hat{e}_{\hat{a}}^{\;\;\mu}$ for a radially free falling observer is then typically chosen so that the timelike vector aligns with the observer's four-velocity. For an observer falling from rest, the tetrad can be written as \cite{Lima:2020wcb,crispim2025}
\begin{eqnarray}
    \hat{e}_{\hat{t}}^{\;\; \mu}&=& \left(\frac{E}{f},\;-\sqrt{E^2 - f},\;0,\;0\right),\\
    \hat{e}_{\hat{r}}^{\;\;\mu}&=& \left(-\frac{\sqrt{E^2 - f}}{f},\;E,\;0,\;0\right),\\
    \hat{e}_{\hat{\theta}}^{\;\;\mu}&=& \left(0,\;0,\;\frac{1}{r},\;0\right),\\
    \hat{e}_{\hat{\varphi}}^{\;\;\mu}&=& \left(0,\;0,\;0,\;\frac{1}{r\sin\theta}\right).
\end{eqnarray}

 These vectors are mutually orthogonal and satisfy the Minkowski normalization condition 
\begin{equation}    \hat{e}_{\hat{a}}^{\;\;\mu}\hat{e}_{\mu}^{\;\;\hat{b}} = \delta_{\hat{a}}^{\;\;\hat{b}},
\end{equation}
ensuring that a local observer measuring the distances and angles in this frame obtains results consistent with special relativity \cite{crispim2025}. Using the tetrad frame, we can compute the components of the Riemann tensor, and consequently, those of the tidal tensor, in the local reference frame defined by the tetrad, where we find
\begin{equation}
    K^{\;\;\hat{b}}_{\hat{a}} = \text{diag}(0,k_1,k_2,k_2),
\end{equation}
where
\begin{equation}
    k_1 = -\frac{ f''}{2 },\quad
    k_2 = -\frac{f'}{2r}.
\end{equation}

With this, we can write the equations that describe the relative acceleration between two nearby observers as
\begin{eqnarray}
\label{tidal1}    \frac{d^2\xi^{\hat{r}}}{d\tau^2} &=& -\frac{ f''}{2 }\xi^{\hat{r}},\\
  \label{tidal2}  \frac{d^2\xi^{\hat{\theta}}}{d\tau^2} &=& -\frac{f'}{2r}\xi^{\hat{\theta}},\\
   \frac{d^2\xi^{\hat{\varphi}}}{d\tau^2} &=& -\frac{f'}{2r}\xi^{\hat{\varphi}}.
\end{eqnarray}

Once the form of $f(r)$ is known, as given by \eqref{geova_solution}, it is straightforward to compute the components of the tidal force, which are given by
\begin{eqnarray}
    k_1&=&\frac{2 M}{r^3}-\frac{2 g_s^2
   l_s^2 \sqrt{\frac{r^4}{l_s^4}+1}}{r^4}+\frac{g_s^2}{l_s^2 \sqrt{\frac{r^4}{l_s^4}+1}}\nonumber\\
   &-&\frac{g_s^2 l_s^2}{r^4} \, _2F_1\left(-\frac{1}{2},-\frac{1}{4};\frac{3}{4};-\frac{r^4}{l_s^4}\right),\\
   k_2&=&-\frac{M}{r^3}+\frac{g_s^2
   l_s^2 \sqrt{\frac{r^4}{l_s^4}+1}}{2 r^4}\nonumber\\
   &+&\frac{g_s^2 l_s^2}{2 r^4} \, _2F_1\left(-\frac{1}{2},-\frac{1}{4};\frac{3}{4};-\frac{r^4}{l_s^4}\right).
\end{eqnarray}
In the limit $g_s \to 0$ (the Schwarzschild case), we obtain
\begin{equation}
    k_1=\frac{2 M}{r^3},\quad \mbox{and} \quad  k_2=-\frac{M}{r^3}.
\end{equation}
Thus, our result is consistent in this limit. Since the difference between the Letelier solution and the Schwarzschild solution is an additive constant in $f(r)$, and the tidal forces depend on the first and second derivatives of $f(r)$, it follows that in the limit $l_s \to 0$ the tidal forces coincide with those of the Schwarzschild case.

\begin{figure*}[htb]
    \centering
    \includegraphics[width=.5\linewidth]{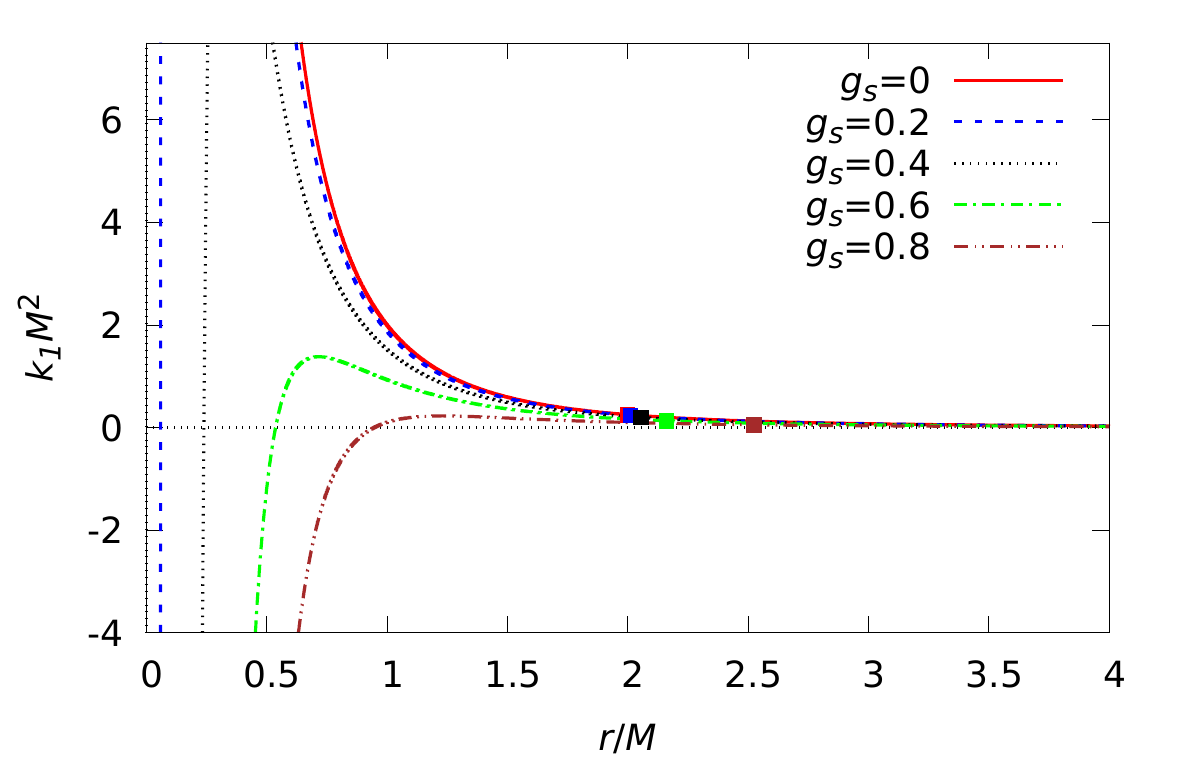}\hspace{-0.1cm}
    \includegraphics[width=.5\linewidth]{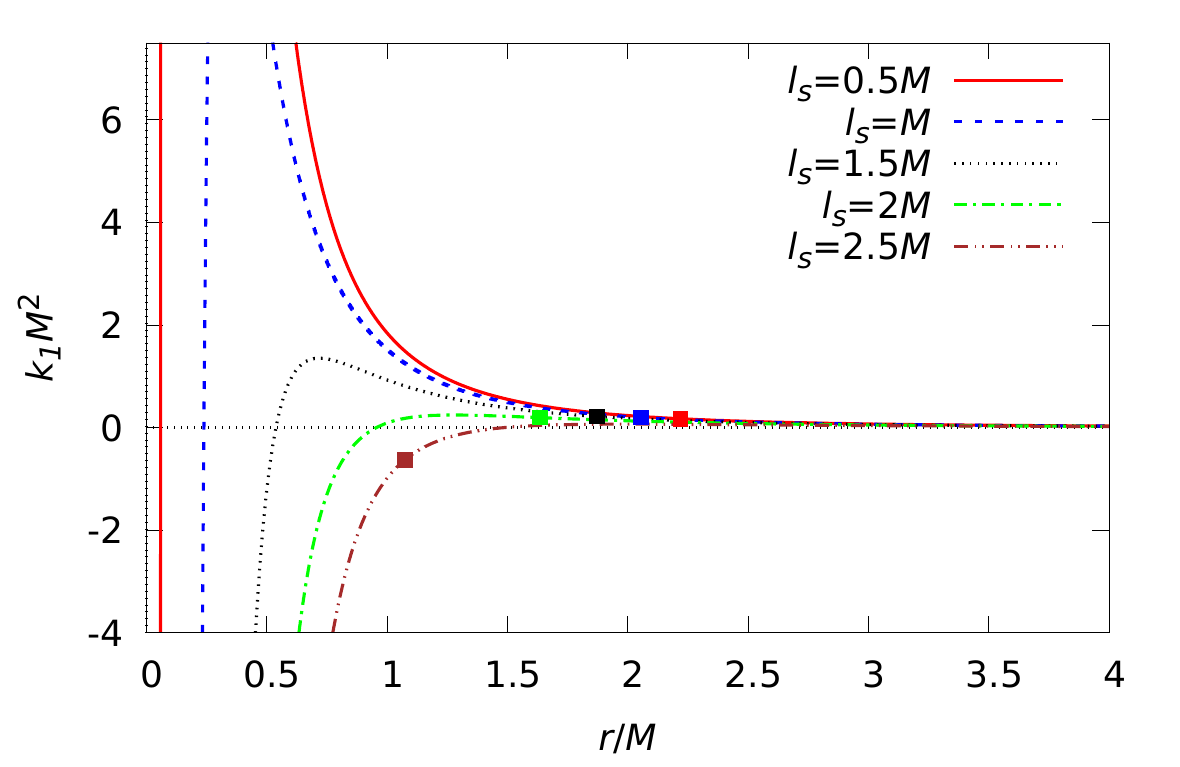}
    \caption{Behavior of the radial component of the tidal force for a frame falling radially into the BH. In the left panel, we set $l_s = M$ and vary $g_s$, whereas in the right panel, we fix $g_s = 0.4$ and vary $l_s$.}
    \label{fig:k1}
\end{figure*}

\begin{figure*}[htb]
    \centering
    \includegraphics[width=.5\linewidth]{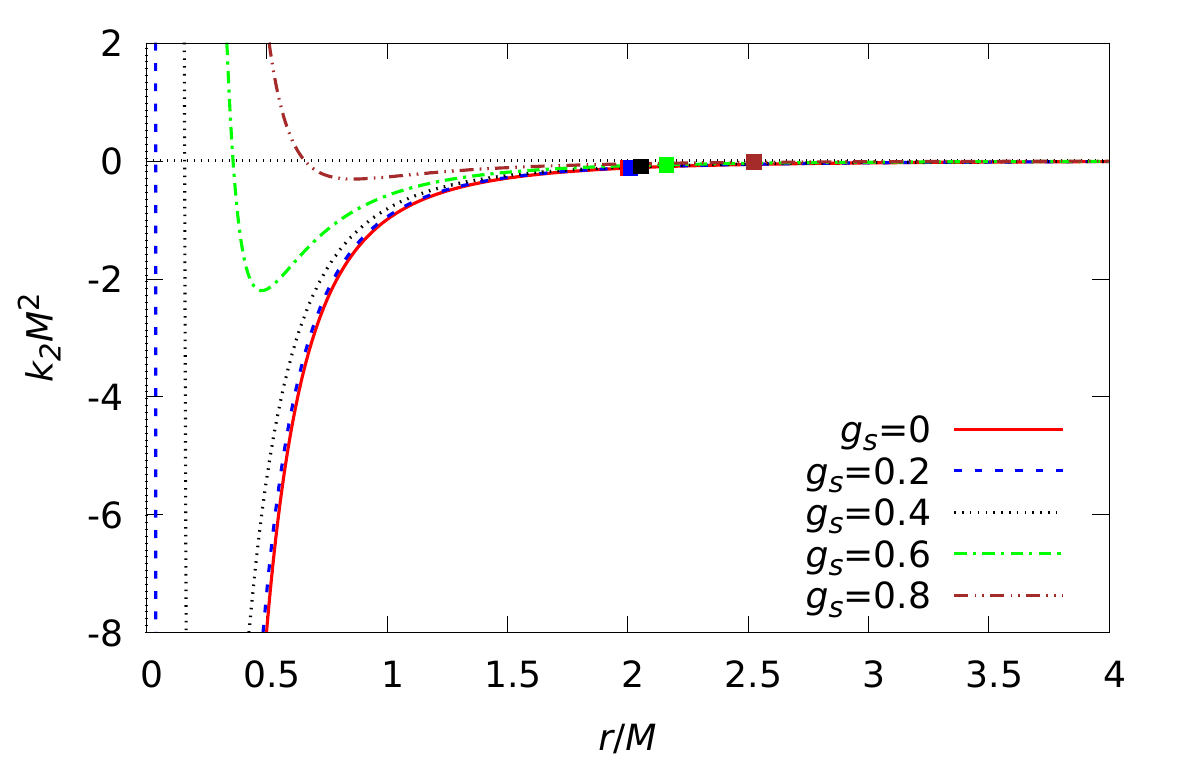}\hspace{-0.1cm}
    \includegraphics[width=.5\linewidth]{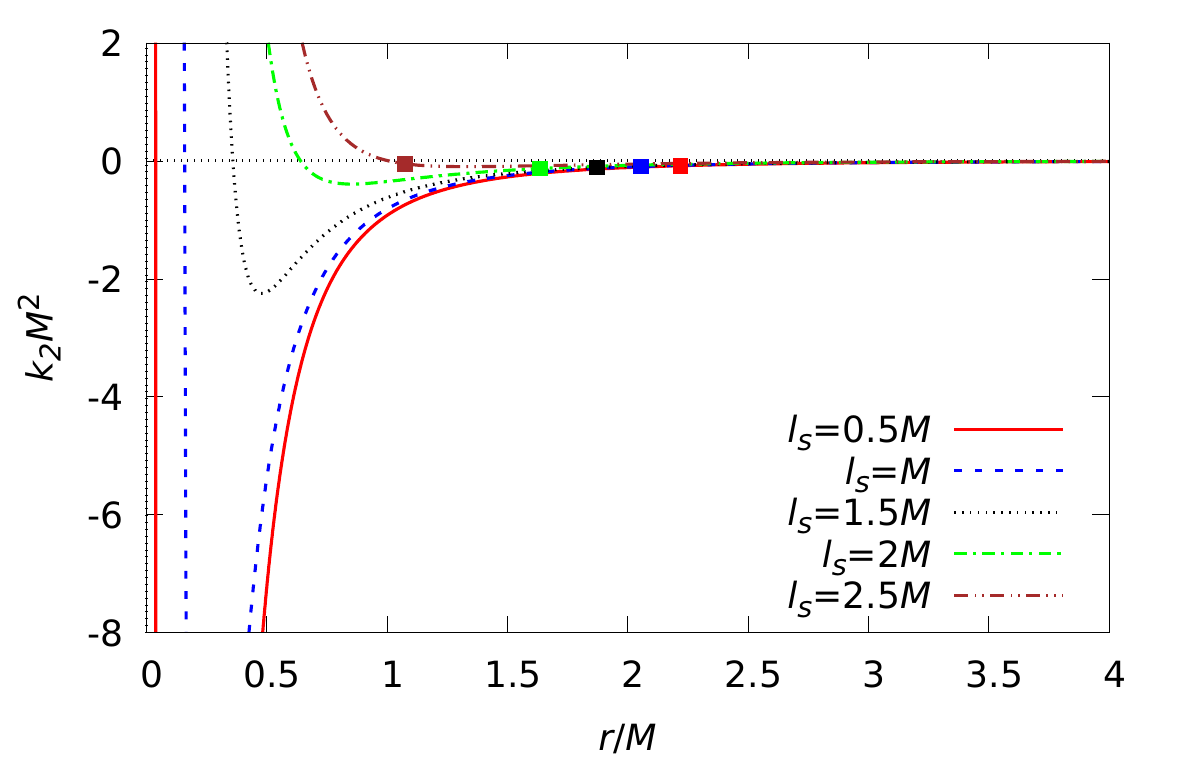}
    \caption{Behavior of the angular component of the tidal force for a frame falling radially into the BH. In the left panel, we set $l_s = M$ and vary $g_s$, whereas in the right panel, we fix $g_s = 0.4$ and vary $l_s$.}
    \label{fig:k2}
\end{figure*}

It is well known that, in the Schwarzschild case, the tidal forces diverge at the singularity. To check whether the same occurs for the generalized cloud of strings solution, we expand the tidal forces for $r \to 0$, which yields
\begin{eqnarray}
    k_1(r\to0) &\approx& -\frac{3 g_s^2 l_s^2}{r^4}+\frac{2 M}{r^3}+\frac{g_s^2}{6 l_s^2}+O\left(r^4\right),\\
    k_2(r\to0) &\approx& \frac{g_s^2 l_s^2}{r^4}-\frac{M}{r^3}+\frac{g_s^2}{6 l_s^2}+O\left(r^4\right).
\end{eqnarray}
Thus, we see that the presence of the cloud of strings causes the tidal forces to diverge even more than in the Schwarzschild case. If we expand the tidal forces for very distant points, we have that
\begin{eqnarray}
    k_1(r\gg M)&\approx&\frac{2 M-\frac{2 g_s^2 l_s \Gamma \left(\frac{3}{4}\right)^2}{\sqrt{\pi }}}{r^3}+O\left(\frac{1}{r^6}\right),\\
    k_2(r\gg M)&\approx&\frac{\frac{g_s^2 l_s \Gamma \left(\frac{3}{4}\right)^2}{\sqrt{\pi }}-M}{r^3}+O\left(\frac{1}{r^6}\right).
\end{eqnarray}
Tidal forces tend to zero at infinity. It is interesting to note that the sign of $k_1$ is negative as $r \to 0$, whereas that of $k_2$ is positive. The sign of the tidal forces at the infinity depends on the combination of $M$, $g_s$, and $l_s$. However, an event horizon does not always exist, depending on this combination of parameters. In Figs.~\ref{fig:k1} and \ref{fig:k2}, we examine the behavior of these components for parameter values that allow the existence of horizons. In all cases shown, a sign reversal occurs in the components of the tidal force, and among the scenarios considered, only for $k_1$ with $g_s=0.4$ and $l_s=2.5M$ does this reversal take place outside the event horizon. In all other cases, the reversal occurs inside the event horizon and thus remains hidden from external observers.

To gain a deeper insight into the tidal forces experienced by the infalling body, we must analyze the evolution of the displacement vector. While the equations of motion are naturally derived in terms of the proper time $\tau$, it is physically more instructive to express the behavior of $\xi^{\mu}$ as a function of the radial distance $r$, allowing us to track the tidal deformations as the body approaches the central object. To achieve this, we perform a change of variables from $\tau$ to $r$. Using the Eq. \eqref{rgeodesica} and chain rule, we can write the following differential equations for the radial ($\xi^{\hat{r}}$) and angular ($\xi^{\hat{i}}$) components of the displacement vector:
\begin{eqnarray}
\left(E^2 - f\right)\xi^{\hat{r}''} - \frac{f'}{2}\xi^{\hat{r}'} + \frac{f''}{2}\xi^{\hat{r}} &=& 0, \label{eq:xi_radial} \\
\left(E^2 - f\right)\xi^{\hat{i}''} - \frac{f'}{2}\xi^{\hat{i}'} + \frac{f'}{2r}\xi^{\hat{i}} &=& 0. \label{eq:xi_angular}
\end{eqnarray}

The general solutions for the radial and transversal components of the displacement vector, derived from the integration of the system \eqref{eq:xi_radial} and \eqref{eq:xi_angular}, can be expressed as \cite{Lima:2020wcb,Crispino:2016pnv}
\begin{eqnarray}
\xi^{\hat{r}}(r) &=& \sqrt{E^2 - f}\left[C_1 + C_2 \int \frac{dr}{\left(E^2 - f\right)^{3/2}}\right], \label{eq:sol_radial} \\
\xi^{\hat{i}}(r) &=& r \left[C_3 + C_4 \int \frac{dr}{r^2 \sqrt{E^2 - f}}\right], \label{eq:sol_angular}
\end{eqnarray}
where the set $\{C_1, C_2, C_3, C_4\}$ represents the integration constants determined by the physical state of the body at the beginning of the fall. Equations \eqref{eq:sol_radial} and \eqref{eq:sol_angular} cannot be expressed in analytic form for the function $f(r)$ given in \eqref{geova_solution}. To proceed with the analysis, we assume the infall starts at a radial coordinate $r=b$ (with $b > r_+$) and consider two distinct kinematic configurations, hereafter referred to as (ICI) and (ICII),
\begin{eqnarray}
\text{(ICI):}  \quad \xi^{\hat{\alpha}}(b) > 0, \quad {\xi^{\hat{\alpha}}}'(b) = 0, \label{eq:cond_ICI} \\
\text{(ICII):} \quad \xi^{\hat{\alpha}}(b) = 0, \quad {\xi^{\hat{\alpha}}}'(b) > 0, \label{eq:cond_ICII}
\end{eqnarray}
where $\hat{\xi}^{\alpha}(b)$ is the $\hat{\alpha} = \{\hat{r}, \hat{\theta}, \hat{\varphi}\}$ component of the infinitesimal displacement vector at $r = b$.
Physically, the first scenario (ICI) describes a cloud of dust released from rest with no initial internal deformation rate. In contrast, the second scenario (ICII) models a dynamic event where the body undergoes an isotropic expansion at the initial radius $r=b$ \cite{Lima:2020wcb,Crispino:2016pnv}.

\begin{figure*}[htb]
    \centering
    \includegraphics[width=.5\linewidth]{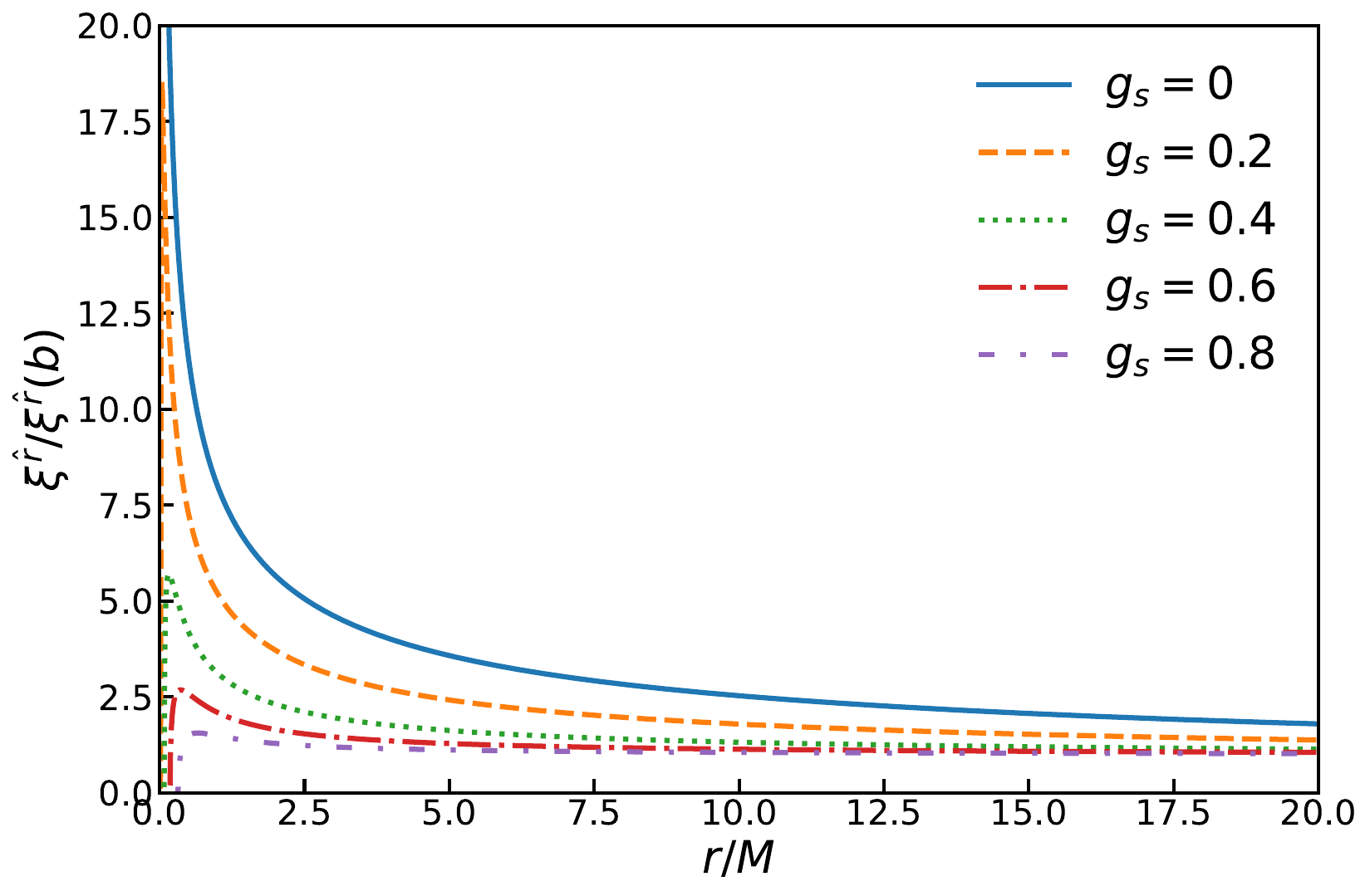}\hspace{-0.1cm}
    \includegraphics[width=.5\linewidth]{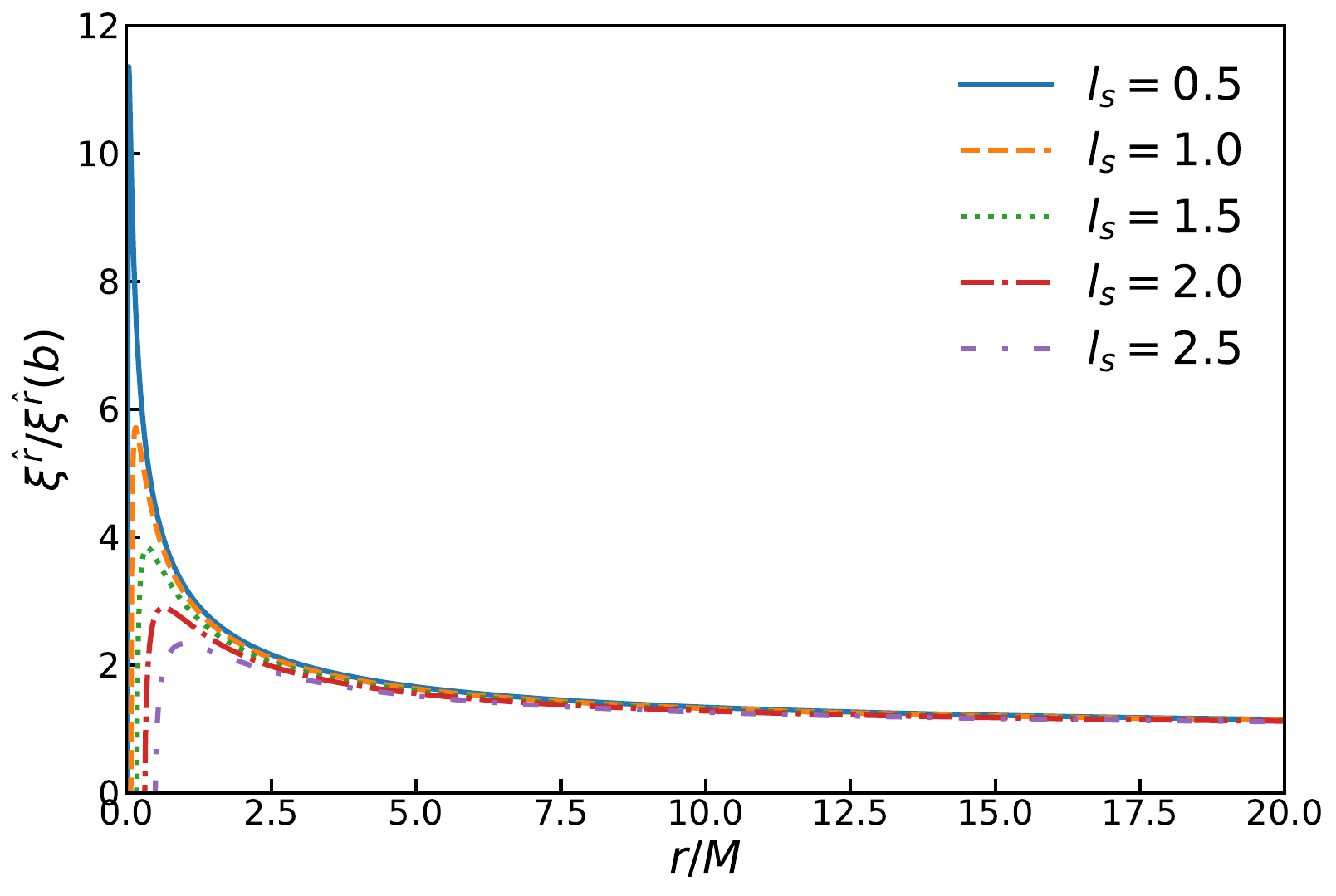}
    \caption{Behavior of the radial component of the displacement vector for a frame falling radially into the BH, with the initial radial position fixed at $b = 100M$ in the (ICI) scenario. In the left panel, we set $l_s = M$ and vary $g_s$, whereas in the right panel we fix $g_s = 0.4$ and vary $l_s$.}
    \label{fig:xir}
\end{figure*}

\begin{figure*}[htb]
    \centering
    \includegraphics[width=.5\linewidth]{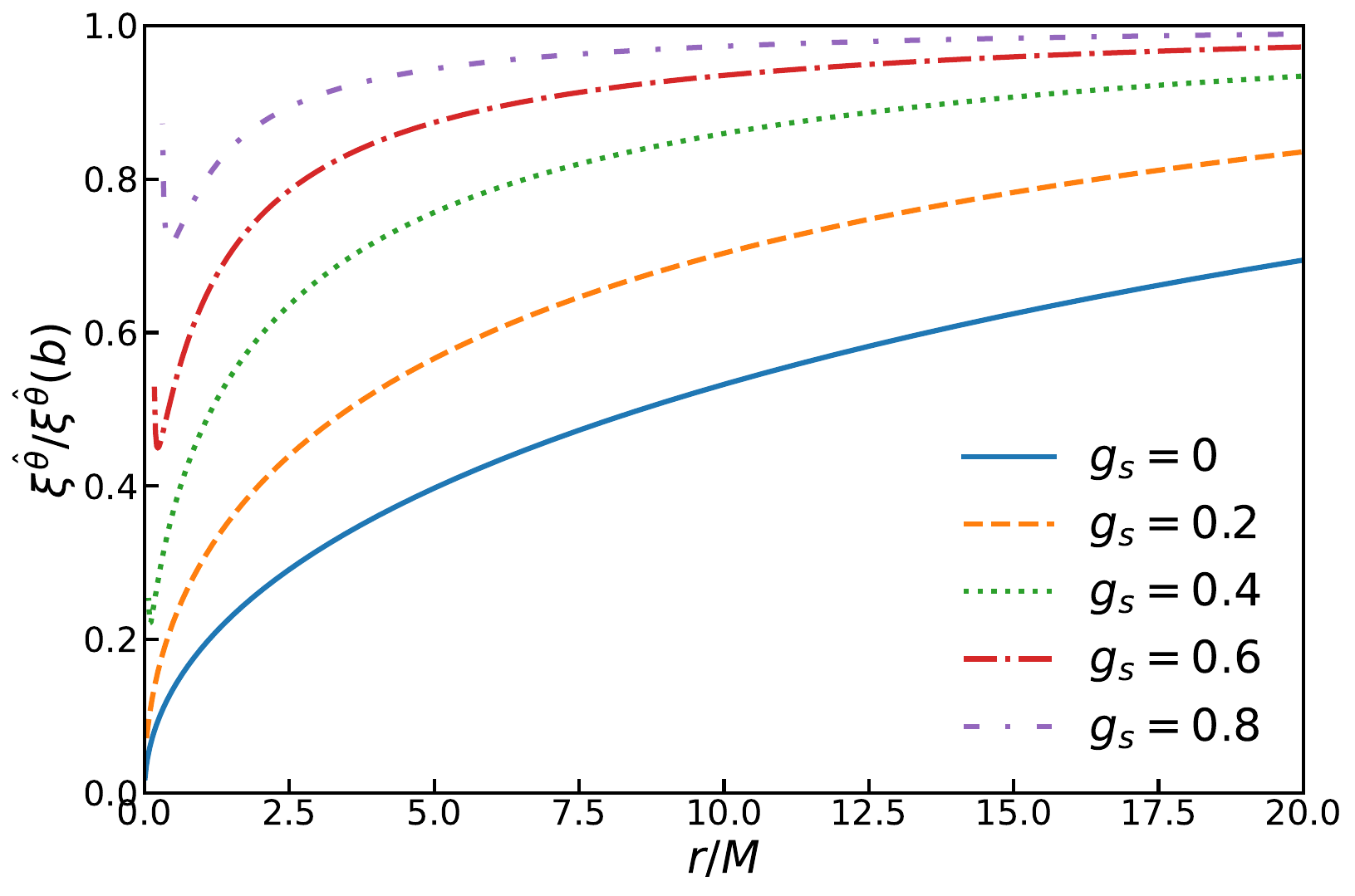}\hspace{-0.1cm}
    \includegraphics[width=.5\linewidth]{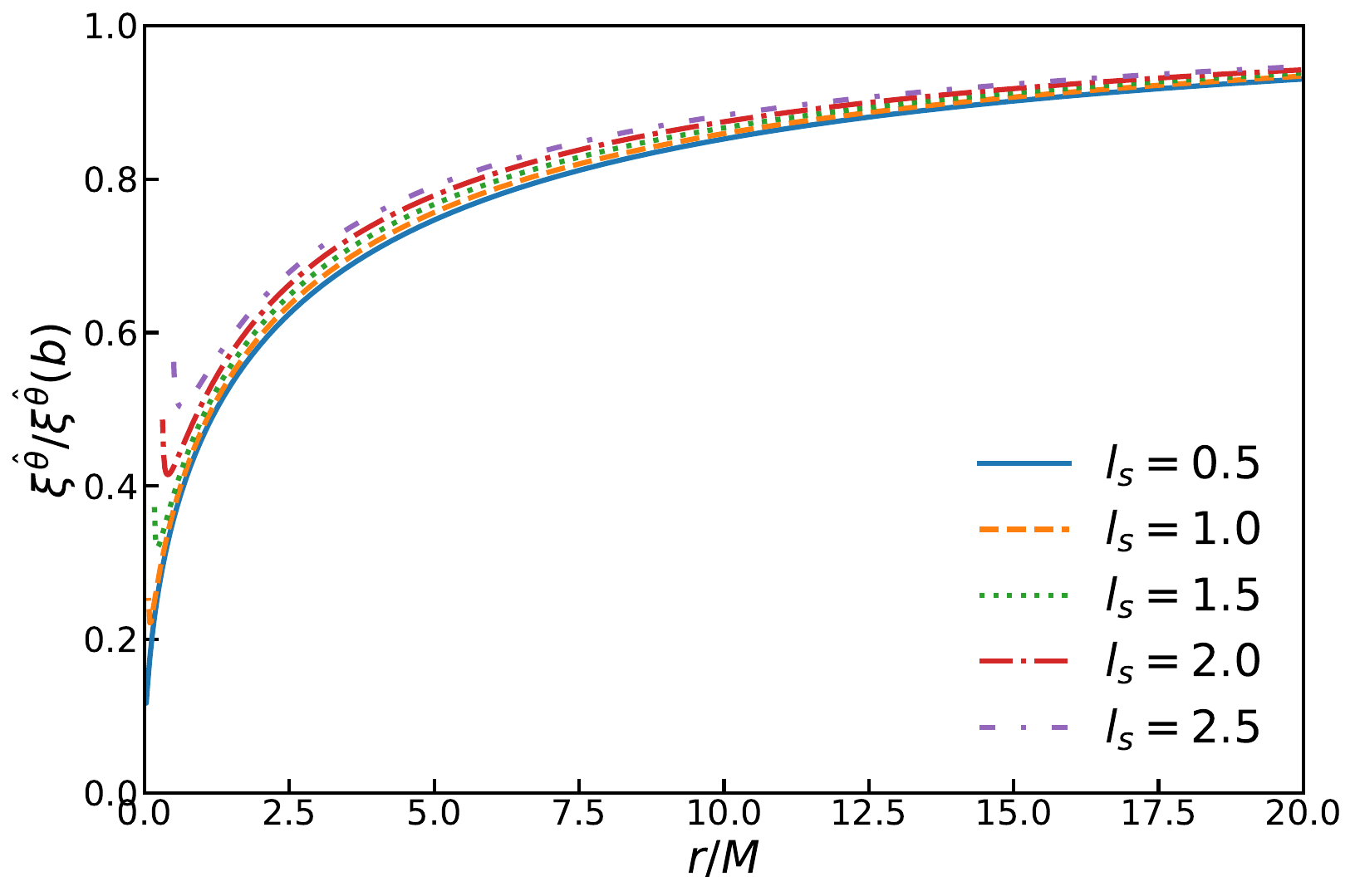}
    \caption{Behavior of the angular component of the displacement vector for a frame falling radially into the BH, with the initial radial position fixed at $b = 100M$ in the (ICI) scenario. In the left panel, we set $l_s = M$ and vary $g_s$, whereas in the right panel we fix $g_s = 0.4$ and vary $l_s$.}
    \label{fig:xia}
\end{figure*}

In Figs.~\ref{fig:xir} and \ref{fig:xia}, we examine the numerical behavior of the components of the displacement vector for parameter values that allow the existence of horizons for the (ICI) scenario. When $g_s = 0$, the stretching behavior of the Schwarzschild case is recovered. For the other parameter values, we observe an inversion of the behavior, such that there are regions where the radial component is compressed instead of stretched, and the angular component is stretched instead of compressed. However, for these parameter values, this behavior is hidden behind the event horizon.  We can highlight that the maximum magnitude of the radial component of the displacement vector decreases as the values of $g_{s}$ and $l_{s}$ increase. Moreover, although its value remains smaller than the initial one, increasing $g_{s}$ and $l_{s}$ allows the angular component of the displacement vector to reach larger values, rather than vanishing as in the Schwarzschild case. In this way, the parameters associated with the cloud of strings reduce the stretching effect as their values increase.

\begin{figure*}[htb]
    \centering
    \includegraphics[width=.5\linewidth]{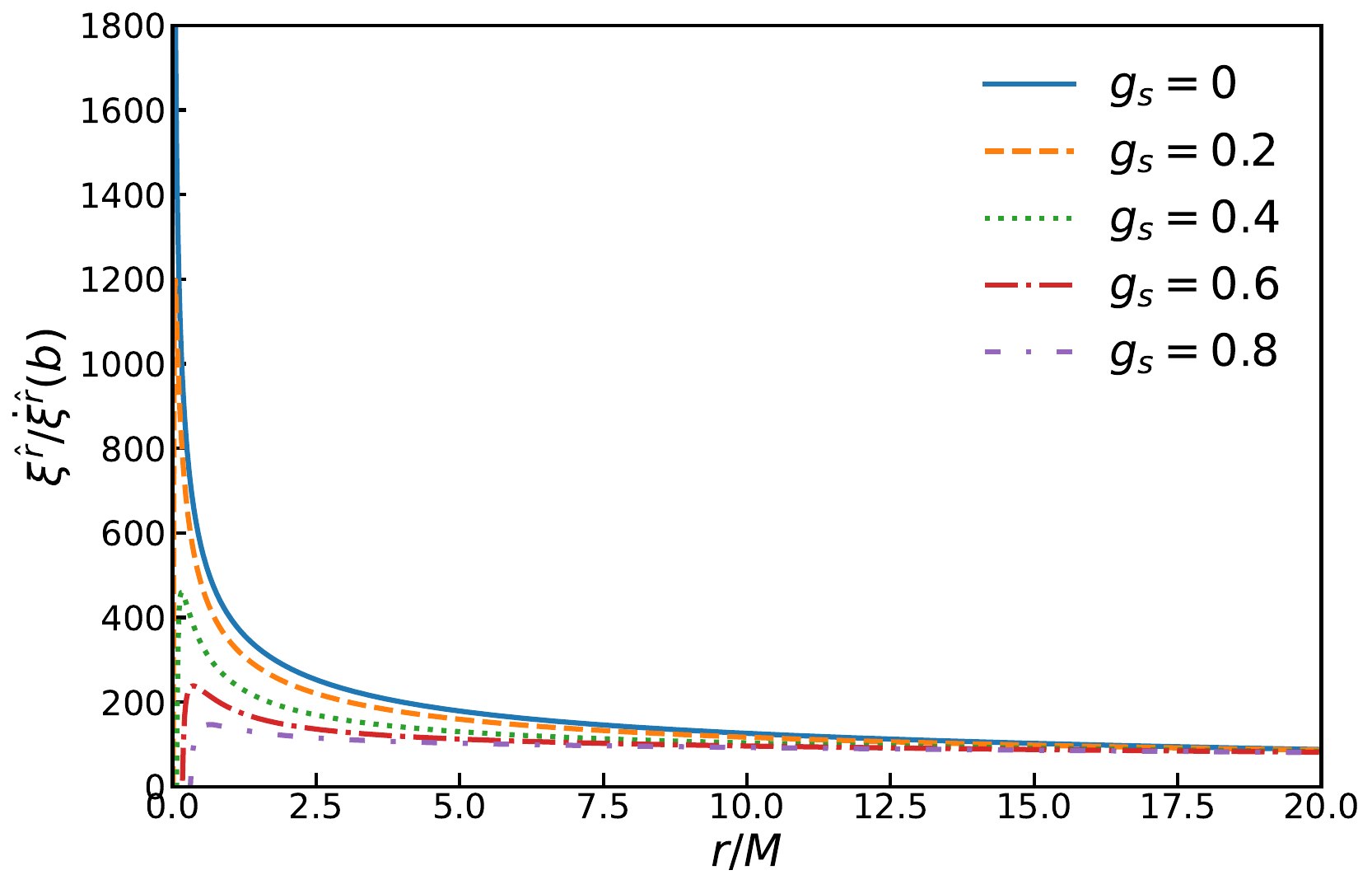}\hspace{-0.1cm}
    \includegraphics[width=.5\linewidth]{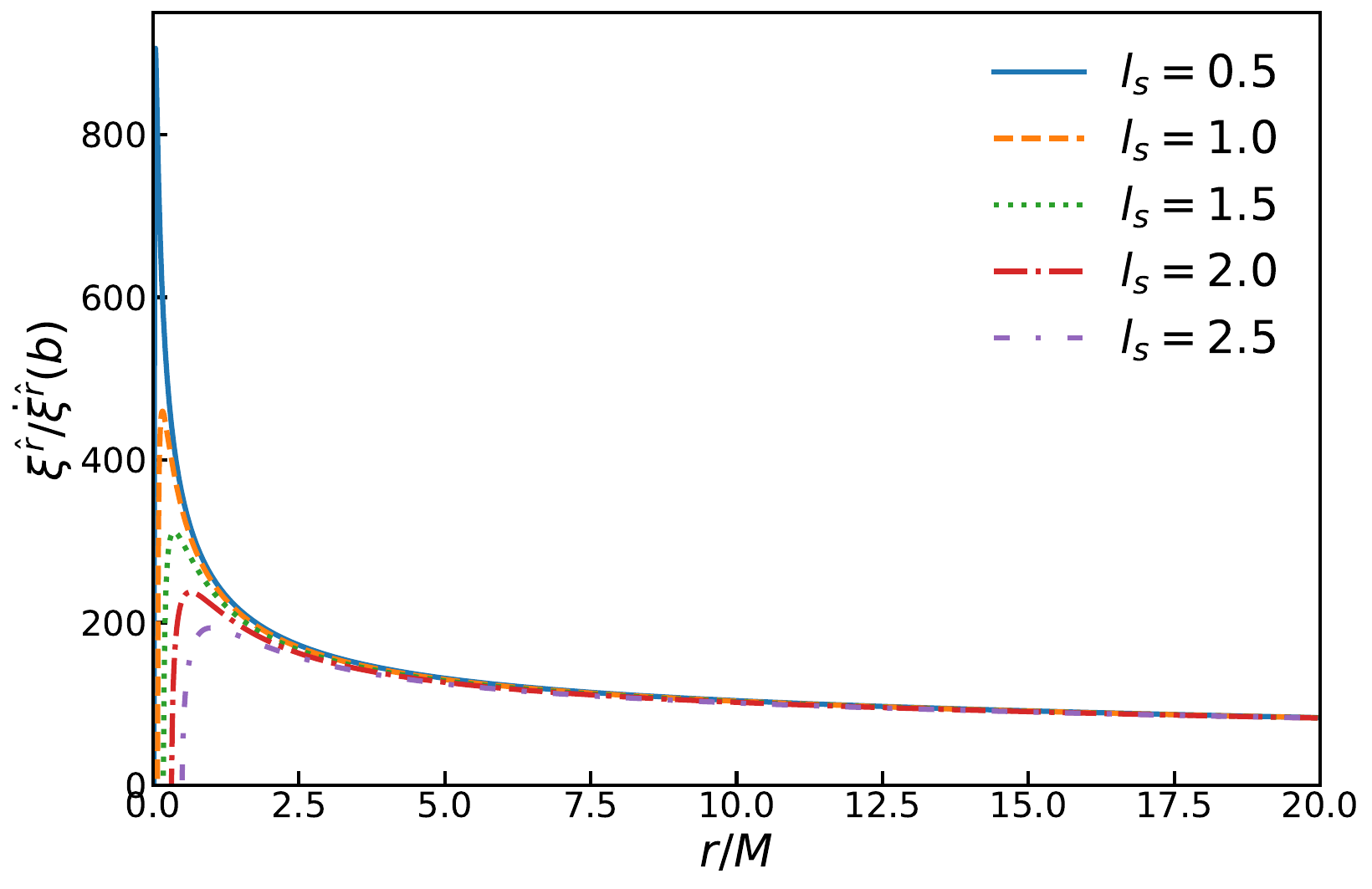}
    \caption{Behavior of the radial component of the displacement vector for a frame falling radially into the BH, with the initial radial position fixed at $b = 100M$ in the (ICII) scenario. In the left panel, we set $l_s = M$ and vary $g_s$, whereas in the right panel we fix $g_s = 0.4$ and vary $l_s$.}
    \label{fig:xirII}
\end{figure*}

\begin{figure*}[htb]
    \centering
    \includegraphics[width=.5\linewidth]{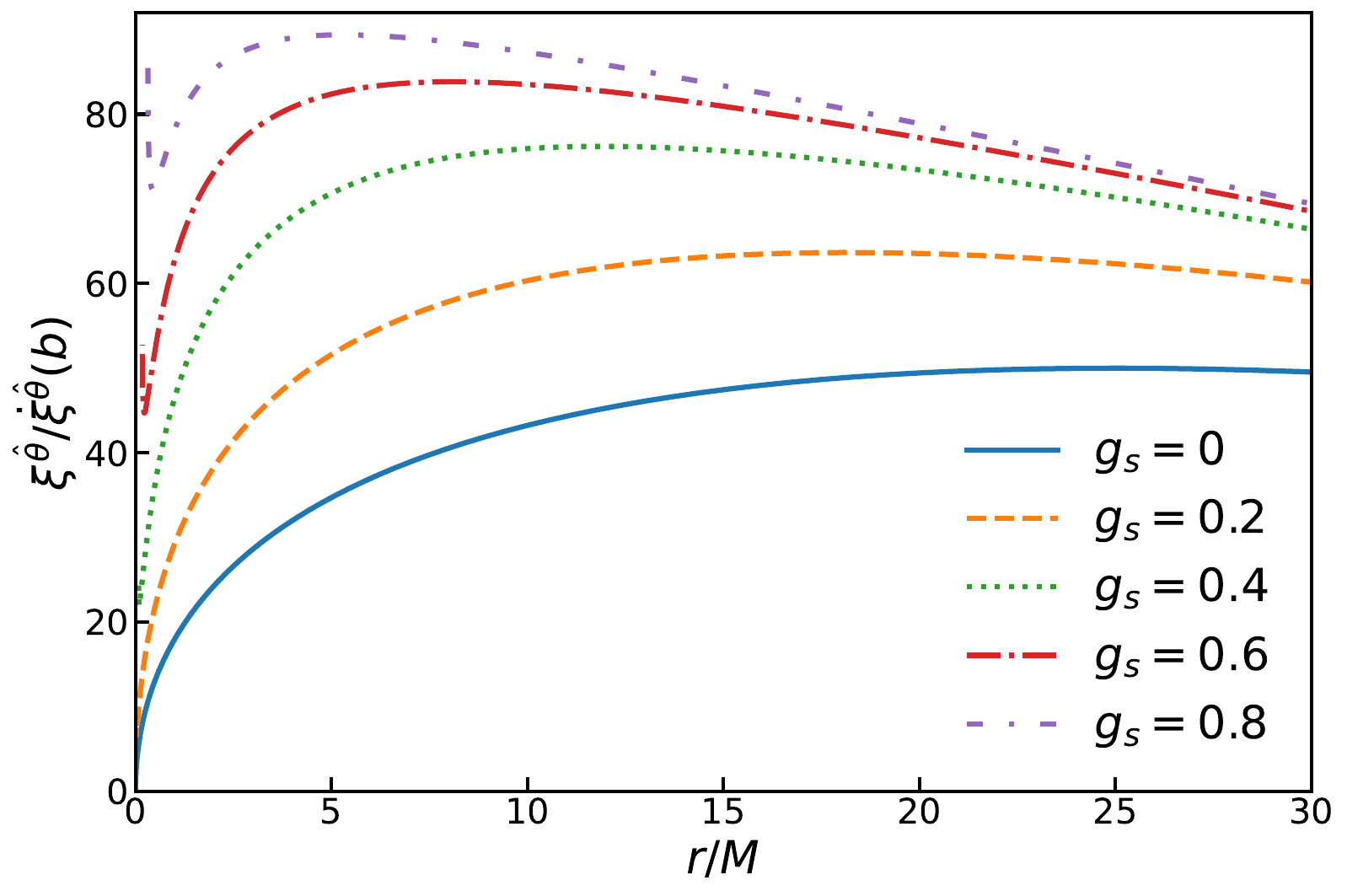}\hspace{-0.1cm}
    \includegraphics[width=.5\linewidth]{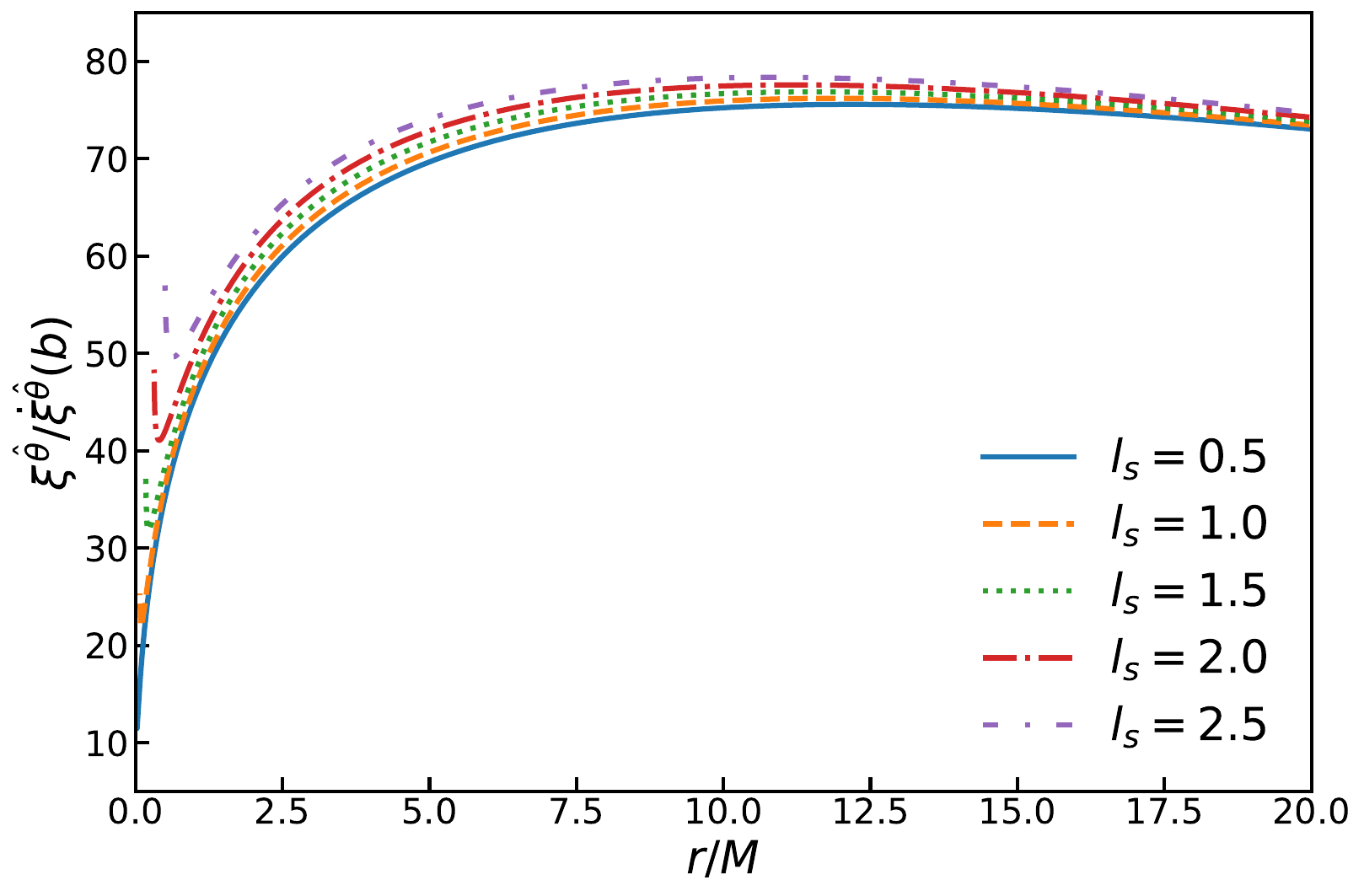}
    \caption{Behavior of the angular component of the displacement vector for a frame falling radially into the BH, with the initial radial position fixed at $b = 100M$ in the (ICII) scenario. In the left panel, we set $l_s = M$ and vary $g_s$, whereas in the right panel we fix $g_s = 0.4$ and vary $l_s$.}
    \label{fig:xiaII}
\end{figure*}

In Figs.~\ref{fig:xirII} and \ref{fig:xiaII}, we examine the numerical behavior of the components of the displacement vector for parameter values that allow the existence of horizons for the (ICII) scenario. From Fig.~\ref{fig:xirII}, we observe that the qualitative behavior of the radial component is quite similar to that of the radial component in the (ICI) scenario, the main difference in this case is that the displacement vector grows much more than in the (ICI) condition before it starts to decrease in magnitude. From Fig.~\ref{fig:xiaII}, we observe that the angular components initially increase, reach a maximum, and then start to decrease as the body falls toward the BH, until reaching a minimum, after which they begin to increase again. This behavior reflects the change in the sign of the tidal forces in the angular directions.

\subsection{Relativistic tidal forces effects in circular geodesics}
In the previously studied case, an observer following a radial geodesic toward the BH has a tetrad basis that is parallel transported along the geodesic, which simplifies writing the geodesic deviation equation in the local frame. In the present case, the observer undergoes a circular motion around the BH, so the natural choice of tetrads is no longer naturally parallel transported \cite{Vieira:2025vwe, Marck:1983mw,Fuchs:1990,vandeMeent:2019cam}. Indeed, given the four-velocity of an orbiting body as
\begin{equation}
    u^\mu = \left(\frac{E_c}{f(r_c)},0,0, \frac{L_c}{r_c^2}\right),
\end{equation}
and choosing again $\hat{e}_{\hat{t}}^{\;\;\mu}$ as the normalized four-velocity, the orthonormal tetrad basis in this case is expressed as \cite{Vieira:2025vwe}
\begin{eqnarray}
\hat{e}_{\hat{t}}^{\;\;\mu} &=& \Delta^{-1/2} \left( 1,0,0,\sqrt{\frac{f-\Delta}{r_c^2}} \right), \\
\hat{e}_{\hat{r}}^{\;\;\mu} &=& \left( 0,\sqrt{f},0,0 \right), \\
\hat{e}_{\hat{\theta}}^{\;\;\mu} &=&  \left( 0,0,\frac{1}{r_c},0 \right), \\
\hat{e}_{\hat{\varphi}}^{\;\;\mu} &=& \Delta^{-1/2} \left( \sqrt{1-\frac{\Delta}{f}},0,0,\frac{\sqrt{f}}{r_c} \right), 
\end{eqnarray}
where $\Delta = f - \frac{1}{2}r_cf'$ and the function $f$ and its derivatives are evaluated at $r=r_c$. Compared with the expression \eqref{LR_eq}, we see that $\Delta \to 0$ when $r_c \to r_{LR}$. 

As discussed previously, in the case where the observer is in circular motion, the tetrad basis attached to the observer is no longer constant, so the projection of equation \eqref{deviation} into the local frame is no longer trivial. Indeed, by writing the components of the infinitesimal displacement vector as $\xi^{\hat{a}} = \xi^{\mu}\hat{e}_{\mu}^{\;\;\hat{a}}$, and recalling that the derivative along the geodesic is defined as $D/D\tau = u \cdot \nabla = u^\mu \nabla_\mu$, we have the relation
\begin{equation}
   \dot{\xi}^{\hat{a}} = \hat{e}_\mu^{\;\;\hat{a}}\dot{\xi}^\mu + \xi^\mu u^\lambda{(\Omega_\lambda)}_{\hat{b}}^{\;\;\hat{a}}\, \hat{e}_\mu^{\;\;\hat{b}},
\end{equation}
where we define ${(\Omega_\lambda)}_{\hat{b}}^{\;\;\hat{a}}\, \hat{e}_\mu^{\;\;\hat{b}} = \nabla_\lambda \hat{e}_\mu^{\;\;\hat{a}}$ and ${(\Omega_\lambda)}_{\hat{c}}^{\;\;\hat{a}}$ as the spin connection, defined in terms of the Christoffel connection as
\begin{equation}
{(\Omega_{\lambda})}_{\hat{c}}^{\;\;\hat{a}} = \hat{e}_{\hat{c}}^{\;\;\mu} \left(\partial_\lambda \hat{e}_\mu^{\;\;\hat{a}} - \Gamma^\alpha_{\lambda\mu}\hat{e}_\alpha^{\;\;\hat{a}}\right).
\end{equation}

By repeating this procedure, we are able to write equation \eqref{deviation} in the local frame in the following form:
\begin{equation}\label{local}
    \frac{d^2\xi^{\hat{a}}}{d\tau^2} = P^{\;\;\hat{a}}_{\hat{b}}\frac{d\xi^{\hat{b}}}{d\tau} + Q^{\;\;\hat{a}}_{\hat{b}}\xi^{\hat{b}},
\end{equation}
where
\begin{eqnarray}
    P^{\;\;\hat{a}}_{\hat{b}} &=&  2u^\nu {(\Omega_\nu)}^{\;\;\hat{a}}_{\hat{b}},\\
    Q^{\;\;\hat{a}}_{\hat{b}} &=& K^{\;\;\hat{a}}_{\hat{b}} - u^\nu u^\lambda{(\Omega_\nu)}^{\;\;\hat{d}}_{\hat{b}}\,{(\Omega_\lambda)}^{\;\;\hat{a}}_{\hat{d}}\\\nonumber
    &+&u^\nu u^\alpha \left[\partial_\alpha{(\Omega_\nu)}_{\hat{b}}^{\;\;\hat{a}} - \Gamma^\lambda_{\alpha\nu}{(\Omega_\lambda)}_{\hat{b}}^{\;\;\hat{a}}\right].
\end{eqnarray}
The term $P^{\;\;\hat{a}}_{\hat{b}} $ is associated with the first time derivative of the displacement vector and is therefore related to Coriolis and centrifugal acceleration due to circular motion, while the term originating from $ Q^{\;\;\hat{a}}_{\hat{b}}$ is associated with the displacement vector itself and is linked to tidal forces. This spin connection formalism is quite general and allows us to write the geodesic deviation equation in general local frames, that is, frames not necessarily in parallel transport along the particle's world-line.

With this, we can finally write equation \eqref{local} in components, from which we obtain
\begin{eqnarray}
    \frac{d^2 \xi^{\hat{r}}}{d\tau^2} &=& \sqrt{\frac{2f'}{r_c}}\frac{d\xi^{\hat{\varphi}}}{d\tau} + \left(\frac{f'}{2r_c} + \frac{f'^2 - 2ff''}{4f - 2r_cf'}\right)\xi^{\hat{r}},\\
    \frac{d^2 \xi^{\hat{\theta}}}{d\tau^2} &=& - \left(\frac{f'}{2r_cf - r_c^2f'}\right)\xi^{\hat{\theta}},\\
    \frac{d^2 \xi^{\hat{\varphi}}}{d\tau^2} &=& - \sqrt{\frac{2f'}{r_c}}\frac{d\xi^{\hat{r}}}{d\tau}.
\end{eqnarray}


As a consistency check, we take the limit $g_s \to 0$ (the Schwarzschild case), where we have
\begin{align}
\frac{d^2 \xi^{\hat{r}}}{d\tau^2} &= 
2\sqrt{\frac{M}{r_c^3}}\,\frac{d\xi^{\hat{\varphi}}}{d\tau}
+ \left[\frac{3M(2M - r_c)}{r_c^3(3M - r_c)}\right]\xi^{\hat{r}}, 
\label{eq:tidal_r}\\
\frac{d^2 \xi^{\hat{\theta}}}{d\tau^2} &= 
\frac{M}{r_c^2(3M - r_c)}\xi^{\hat{\theta}}, 
\label{eq:tidal_theta}\\
\frac{d^2 \xi^{\hat{\varphi}}}{d\tau^2} &= 
-2\sqrt{\frac{M}{r_c^3}}\,\frac{d\xi^{\hat{r}}}{d\tau}.
\label{eq:tidal_phi}
\end{align}
 For convenience, we can rewrite the above equations in the following form:
\begin{align}
\frac{d^{2}\xi^{\hat{r}}}{d\tau^{2}} &= 2\,\omega_{K}\,\frac{d\xi^{\hat{\varphi}}}{d\tau}
    + \left(k^{2} + 2\,\omega_{\varphi}^{2}\right)\xi^{\hat{r}}, \label{eq:xi_r_rew}\\
\frac{d^{2}\xi^{\hat{\theta}}}{d\tau^{2}} &= -\,\omega_{\varphi}^{2}\,\xi^{\hat{\theta}}, \label{eq:xi_theta_rew}\\
\frac{d^{2}\xi^{\hat{\varphi}}}{d\tau^{2}} &= -\,2\,\omega_{K}\,
    \frac{d\xi^{\hat{r}}}{d\tau}, \label{eq:xi_phi_rew}
\end{align}
where we have defined the characteristic frequencies as
\begin{equation}
\omega_{\varphi} = \omega_{K}\sqrt{\frac{r_c}{r_c-3M}}, \qquad
k = \omega_{K}\sqrt{\frac{r_c-6M}{r_c-3M}}.
\label{eq:freq_def}
\end{equation}
In the Newtonian limit, \(r \gg M\), the three frequencies coincide,
\(\omega_{K} \simeq \omega_{\varphi} \simeq k\),
and Eqs.~\eqref{eq:xi_r_rew}–\eqref{eq:xi_phi_rew} reduce to the classical
tidal equations for the gravitational field of a central body, Eqs~\eqref{eq:n1}–\eqref{eq:n3}. As expected, these frequencies diverge at the light ring \(r_c = 3M\),
where circular null geodesics are located. In addition, the behavior of
the radial frequency $k$ encodes the stability of circular orbits:
for $r_c > 6M$ the quantity $k^{2} > 0$, corresponding to stable
timelike circular motion, whereas for $r_c < 6M$ one has $k^{2} < 0$,
indicating that small radial perturbations grow exponentially and the
orbit becomes unstable. The radius $r_c = 6M$ therefore marks the
ISCO in the Schwarzschild geometry.

Before considering the full effects of the cloud of strings described by the function \eqref{geova_solution}, it is instructive to analyze the asymptotic behavior \( r_c \gg M \). Considering the limit $r_c \gg M$ with $g_s \neq 0$, we obtain
\begin{eqnarray}
    \frac{d^2 \xi^{\hat{r}}}{d\tau^2} &\approx& 2\sqrt{\frac{M}{r_c^3} - \frac{g_s^2l_s\Gamma(\frac{3}{4})^2}{\sqrt{\pi}r_c^3}}\frac{d\xi^{\hat{\varphi}}}{d\tau}\\\nonumber
    &+& 3\left(\frac{M}{r_c^3} - \frac{g_s^2l_s\Gamma(\frac{3}{4})^2}{\sqrt{\pi}r_c^3}\right)\xi^{\hat{r}},\\
    \frac{d^2 \xi^{\hat{\theta}}}{d\tau^2} &\approx& - \left(\frac{M}{(1 - g_s^2)r_c^3} -\frac{g_s^2l_s\Gamma(\frac{3}{4})^2}{\sqrt{\pi}(1 - g_s^2)r_c^3}\right)\xi^{\hat{\theta}},\\
    \frac{d^2 \xi^{\hat{\varphi}}}{d\tau^2} &\approx& - 2\sqrt{\frac{M}{r_c^3} - \frac{g_s^2l_s\Gamma(\frac{3}{4})^2}{\sqrt{\pi}r_c^3}}\frac{d\xi^{\hat{r}}}{d\tau}.
\end{eqnarray}
In this asymptotic regime, the cloud of strings modifies the tidal dynamics through effective corrections proportional to \( g_s^2 l_s \). 
The effective Keplerian frequency can be read from the first and third equations as
\begin{equation}
{\omega_{\text{eff}}}\,^2 \simeq \frac{M}{r_c^3} - \frac{g_s^2 l_s \Gamma\!\left(\tfrac{3}{4}\right)^2}{\sqrt{\pi}\,r_c^3},
\label{eq:omega_eff}
\end{equation}
which reduces to the standard Newtonian value \(\omega_K^2 = M/r_c^3\) in the Schwarzschild limit \( g_s \to 0 \). 
The presence of the term proportional to $g_s^2 l_s$ indicates that the cloud of strings effectively weakens the gravitational tidal field, leading to smaller oscillation frequencies for the same orbital radius. Moreover, to obtain ${\omega_{\text{eff}}}^2>0$ in this asymptotic limit, it is necessary that $M > g_s^2l_s\Gamma(\frac{3}{4})^2/\sqrt{\pi}$, highlighting that, even in the asymptotic regime, the presence of the cloud of strings plays a decisive role in the stability of the orbits.

The angular and radial equations can then be expressed, in analogy with the Newtonian system, as
\begin{align}
\frac{d^{2}\xi^{\hat{r}}}{d\tau^{2}} &\approx 2\,\omega_{\text{eff}}\frac{d\xi^{\hat{\varphi}}}{d\tau}
+ 3\,\omega_{\text{eff}}^{2}\,\xi^{\hat{r}}, \label{eq:xi_r_string}\\
\frac{d^{2}\xi^{\hat{\theta}}}{d\tau^{2}} &\approx -\,\frac{\omega_{\text{eff}}^{2}}{(1 - g_s^{2})}\,\xi^{\hat{\theta}}, \label{eq:xi_theta_string}\\
\frac{d^{2}\xi^{\hat{\varphi}}}{d\tau^{2}} &\approx -\,2\,\omega_{\text{eff}}\frac{d\xi^{\hat{r}}}{d\tau}.
\label{eq:xi_phi_string}
\end{align}
Therefore, even in the weak-field limit, the presence of the generalized cloud of strings introduces anisotropic tidal corrections, i.e, 
while the in-plane motion (\(\xi^{\hat{r}}, \xi^{\hat{\varphi}}\)) retains the same coupling structure as in the Schwarzschild case, 
the vertical motion (\(\xi^{\hat{\theta}}\)) becomes rescaled by a factor \((1 - g_s^{2})^{-1}\). 
This suggests that the string parameter \( g_s \) not only modifies the strength of the gravitational coupling but also induces a small deviation from isotropy in the local tidal field.

In particular, expanding for small \( g_s \ll 1 \), we find to leading order:
\begin{equation}
\omega_{\text{eff}} \simeq \omega_K\left(1 - \frac{1}{2}\frac{g_s^2 l_s \Gamma\!\left(\tfrac{3}{4}\right)^2}{\sqrt{\pi}\,M}\right),
\end{equation}
and
\begin{equation}
\omega_{\theta} \simeq \omega_K\left(1 + \frac{1}{2}g_s^2\right),
\end{equation}
showing that the radial and angular frequencies no longer coincide, that its, a signature of the effective ``tidal anisotropy'' induced by the cloud of strings. 
This splitting between \(\omega_{\text{eff}}\) and \(\omega_{\theta}\) could, in principle, lead to precession effects or resonance phenomena in circular orbits \cite{Kato2011,AbramowiczKluzniak2003}, under the influence of the string medium.

Even in the standard Letelier cloud of strings limit (\(l_s = 0\)), 
the tidal dynamics remain anisotropic due to the presence of the 
string parameter \(g_s\). 
In this case (where $\omega_{\text{eff}} = \omega_K$), the effective equations for the perturbations still yield 
a distinct polar component of the form 
\begin{equation}
    \frac{d^{2}\xi^{\hat{\theta}}}{d\tau^{2}} \simeq 
-\frac{\omega_{K}^{2}}{(1 - g_s^{2})}\,\xi^{\hat{\theta}},
\end{equation}
while the in-plane (radial and azimuthal) oscillations keep the 
Keplerian-like coupling structure. 
This factor \( (1 - g_s^{2})^{-1} \) originates directly from the 
anisotropic stress-energy tensor of the Letelier cloud of string, 
which modifies only the angular part of the local curvature.
Consequently, the Letelier background itself introduces a 
direction-dependent (anisotropic) tidal response, 
even in the absence of additional corrections 
proportional to \(l_s\).
Such anisotropy implies that the vertical frequency 
\(\omega_{\theta}\) differs from the Keplerian frequency
\(\omega_{K}\), potentially leading to precessional or resonant effects 
in circular motion even in this simple case.

\begin{figure*}[htb]
    \centering
    \includegraphics[width=.5\linewidth]{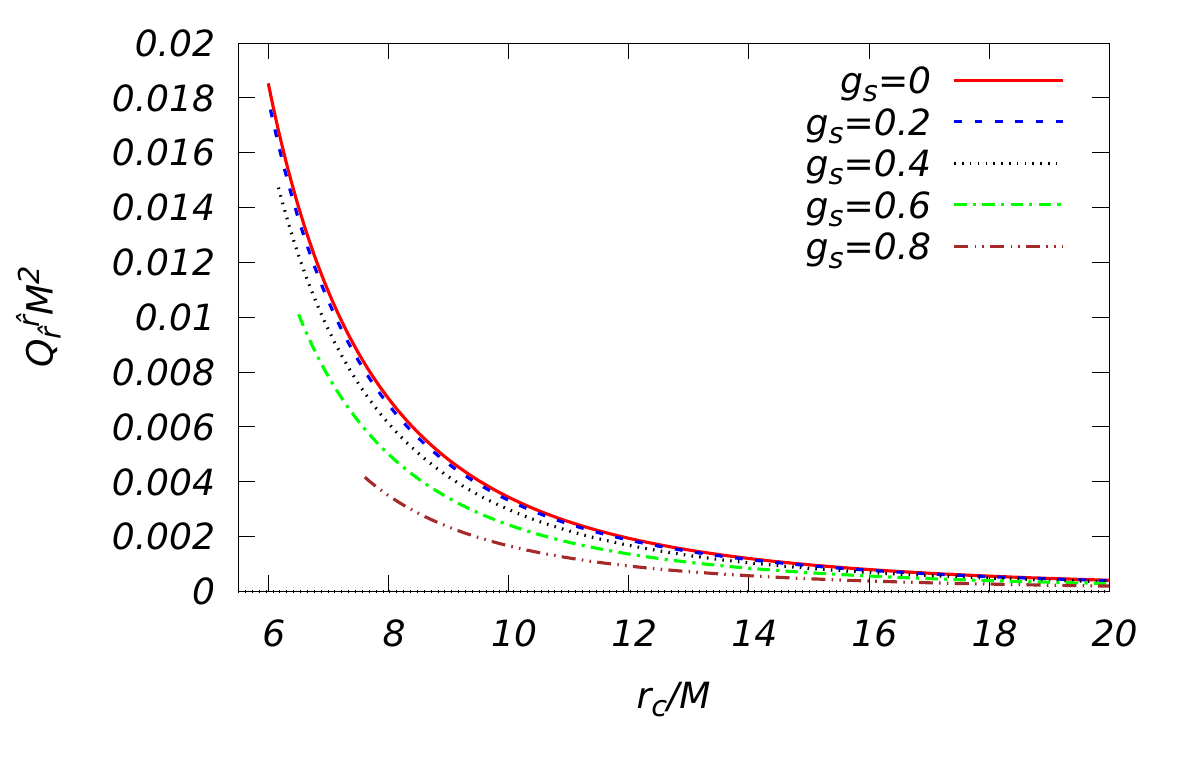}\hspace{-0.1cm}
    \includegraphics[width=.5\linewidth]{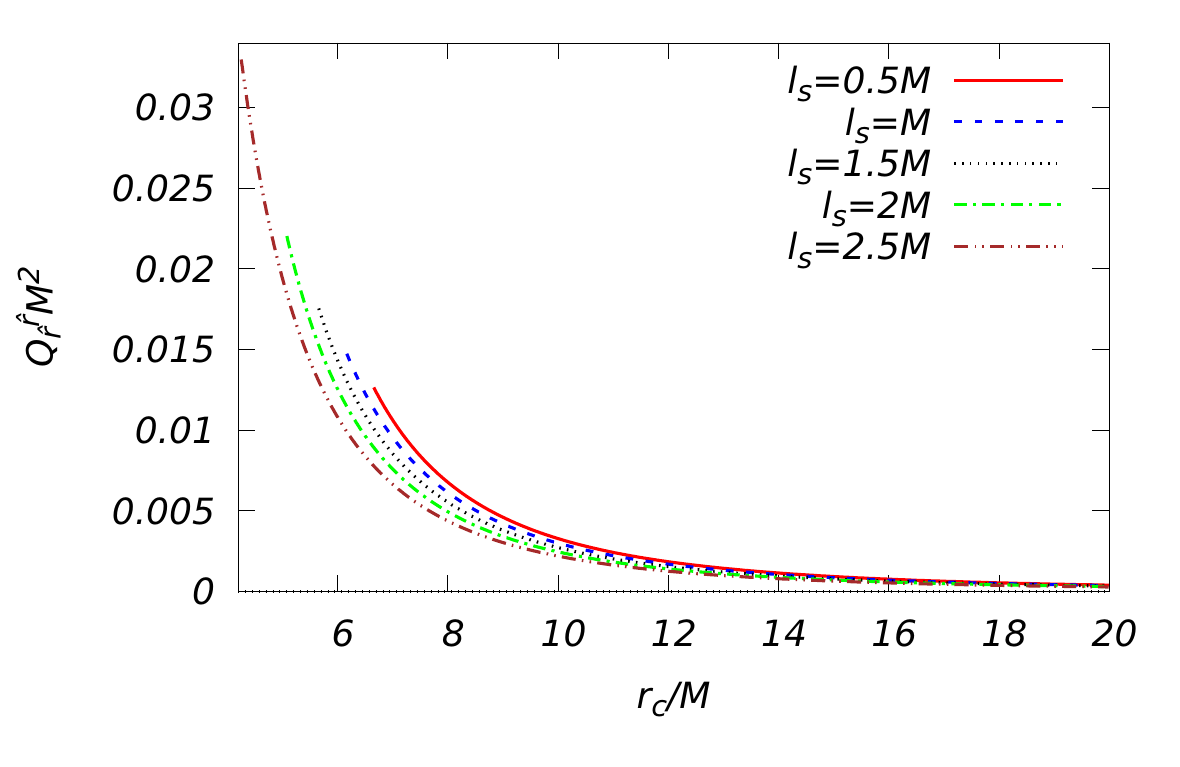}
    \caption{Behavior of the radial component of the tidal force  acting on a body in a circular orbit around the BH. In the left panel, we set $l_s = M$ and vary $g_s$, whereas in the right panel, we fix $g_s = 0.4$ and vary $l_s$. We plot only up to the point where the radius of the circular orbit equals the ISCO radius.}
    \label{fig:k1c}
\end{figure*}

\begin{figure*}[htb]
    \centering
    \includegraphics[width=.5\linewidth]{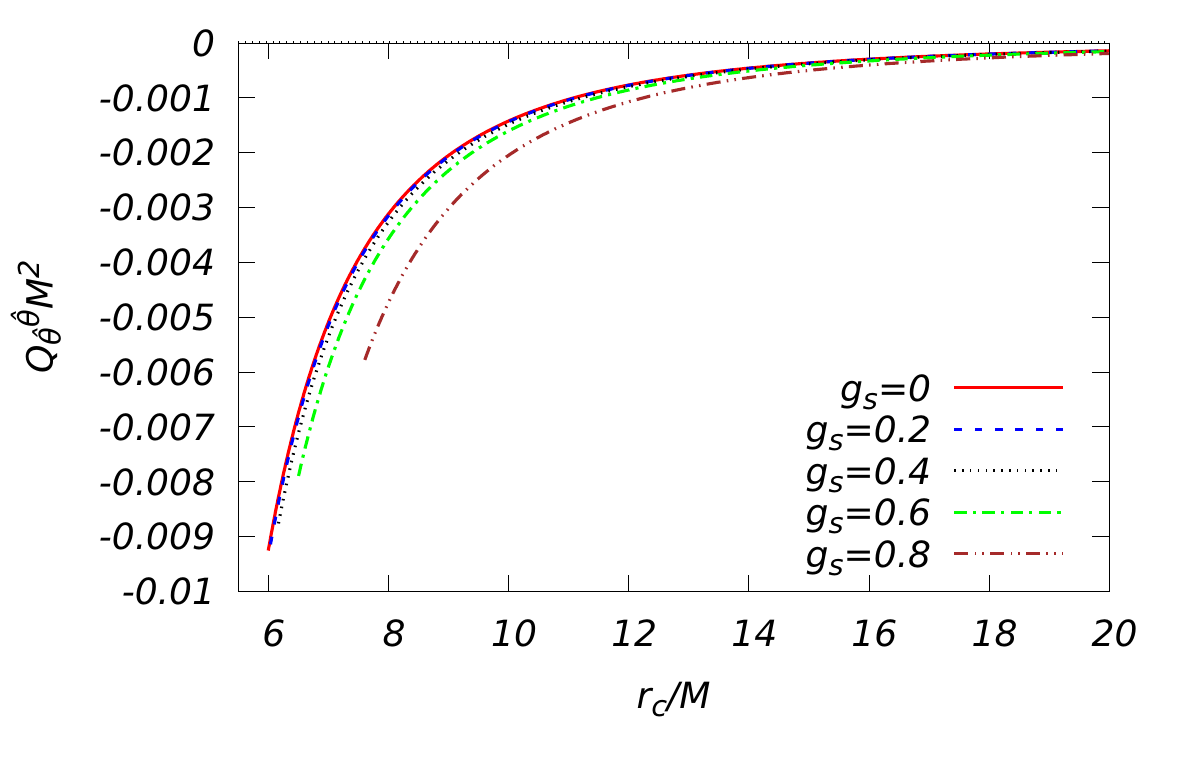}\hspace{-0.1cm}
    \includegraphics[width=.5\linewidth]{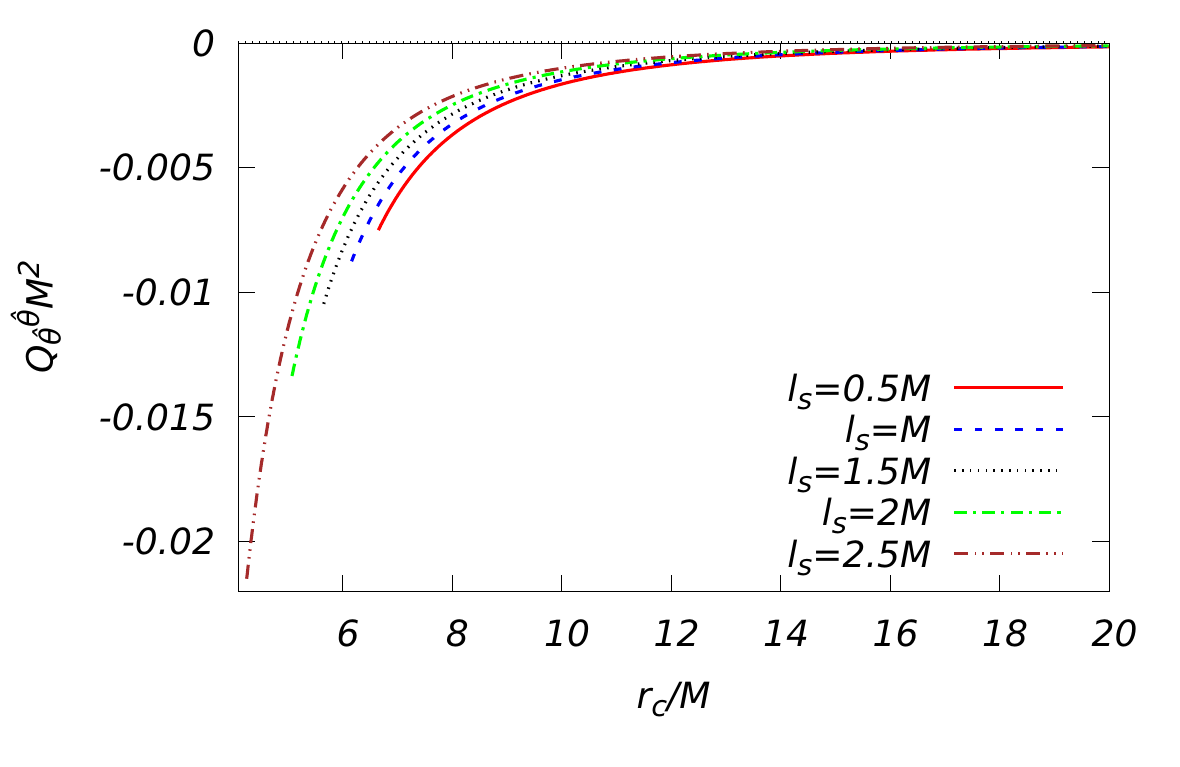}
    \caption{Behavior of the angular component of the tidal force acting on a body in a circular orbit around the BH. In the left panel, we set $l_s = M$ and vary $g_s$, whereas in the right panel, we fix $g_s = 0.4$ and vary $l_s$. We plot only up to the point where the radius of the circular orbit equals the ISCO radius.}
    \label{fig:k2c}
\end{figure*}

If we focus solely on the tidal forces for the Letelier-Alencar solution, we must examine the components of ${Q_{\hat{b}}}^{\hat{a}}$ for this model. In Fig.~\ref{fig:k1c}, we show the behavior of the radial component for different values of the parameters associated with the cloud of strings. For the same circular orbit radius, the radial component of the tidal force decreases in magnitude as $g_s$ increases. Moreover, for larger $g_s$, the radius of the ISCO also increases, placing circular orbits farther out and further reducing the magnitude of the radial component. When we vary $l_s$, we find that, for a fixed radius of the orbit, the magnitude of the radial component decreases. However, a larger $l_s$ allows for smaller orbit radii, and in regions closer to the center, the radial component is more intense.

In Fig.~\ref{fig:k2c}, we show the behavior of the angular component for different values of the parameters associated with the cloud of strings. For the same circular orbit radius, the angular component of the tidal force increases in magnitude as $g_s$ increases. However, for larger $g_s$ the ISCO radius also increases, placing circular orbits farther out, which reduces the magnitude of the angular component; thus, smaller $g_s$ can reach higher intensities of this component. When we vary $l_s$, we find that, for a fixed orbit radius, the magnitude of the angular component decreases as $l_s$ increases. Nevertheless, a larger $l_s$ allows for smaller orbit radii, and in regions closer to the center the angular component becomes more intense.

In the analysis above, we refer to the magnitude of the tidal forces, i.e., the absolute value of those forces. There is no sign change, so the stretching regime does not turn into compression. We also consider the effect of tidal forces only on bodies in stable circular orbits and, therefore, take $r_c \geq r_I$. For $r_c < r_I$, only unstable orbits occur; therefore, it would not be meaningful to analyze tidal forces in such systems.

\section{Conclusions}\label{SEC:conclusions}

In this work, we investigated the effects of tidal forces in the Letelier–Alencar cloud of strings spacetime. Although it is a singular spacetime, much like the Schwarzschild geometry, this metric stands out because the presence of the cloud of strings introduces a nontrivial modification of the curvature structure, which can be physically interpreted as the imprint of the string tension distributed throughout the spacetime. This additional contribution effectively changes the balance between attractive and repulsive tidal components, leading to a richer geodesic deviation behavior compared to the standard Schwarzschild case.  

The causal structure of this geometry is strongly influenced by the parameters $g_s$ and $l_s$. Depending on their values, the spacetime may exhibit two horizons, an event horizon and a Cauchy horizon, an extremal horizon, or even no horizon at all, leading to a naked singularity. This rich causal structure induced by the cloud of strings profoundly affects the geodesic motion. As we have shown, it has a direct and significant impact on the tidal dynamics experienced along geodesics.

Before that, however, we carried out a detailed study of the geodesic motion of massive particles and photons in this spacetime. In doing so, it is interesting to note that, for photons, we showed that an unstable circular orbit always exists, whose position is directly influenced by the combination of the parameters $g_s$ and $l_s$. Although it is not possible to obtain its exact location analytically for the full metric function \eqref{geova_solution}, we demonstrated its existence numerically. Our results indicate that the radius of this unstable orbit reaches its maximum value in the limit $l_s \to 0$, corresponding to the Schwarzschild case, and decreases as $l_s$ increases. This behavior reveals that the presence of the string medium effectively strengthens the gravitational potential, causing the photon sphere to move inward. Consequently, the propagation of null geodesics and the resulting optical properties of the spacetime become sensitive probes of the underlying string distribution.

For massive particles, we analyzed both radial and circular motion and found that the ISCO always exists, with its position directly affected by the parameters of the cloud of strings. As $g_s$ increases, the orbit moves closer to the BH, while a minimum value of $l_s$ is required for its existence. Furthermore, in the asymptotic limit $r \gg M$, the effective potential tends to the nonzero constant $(1 - g_s^2)$, showing that the influence of the cloud of strings extends to large distances and affects the geodesic dynamics even in the asymptotic regime.

To finally analyze the tidal force effects in this spacetime, we employed the tetrad formalism, which allowed us to express the geodesic deviation equation in the observer’s local frame. Our analysis was divided into two main parts. In the first one, we examined the tidal effects on a radial freely falling observer. Our results show that, in the Letelier limit ($l_s = 0$), the tidal forces coincide exactly with those of the Schwarzschild case, since the Letelier solution differs from the Schwarzschild metric only by an additive constant in the gravitational potential. When we consider the full Letelier-Alencar solution, we found that the presence of the cloud of strings enhances the tidal forces compared to the Schwarzschild case. This behavior arises from the contribution of the string tension to the curvature, which increases the anisotropy of the spacetime and amplifies the relative acceleration between nearby geodesics. In the Letelier-Alencar spacetime, depending on the values of the parameters $g_s$ and $l_s$, the tidal components may even change sign, indicating a local inversion between stretching and compression. Such inversions, when occurring outside the event horizon, could, in principle lead to observable differences in the tidal environment of compact objects.

We computed the displacement vector in order to determine how the modifications in the tidal forces can affect the stretching experienced by extended bodies. For this purpose, we considered two types of boundary conditions commonly adopted in the literature. In the (ICI) condition, the body is released from rest and falls radially into the black hole. We found that the radial component of the displacement vector increases until it reaches a maximum value and then decreases, eventually vanishing. This behavior differs from the Schwarzschild case, where the radial component grows without bound and diverges at the black hole singularity. The maximum value of the radial component decreases as the cloud of strings parameters $g_{s}$ and $l_{s}$ increase, thereby reducing the stretching effect experienced by the body. For the angular component, we found that the displacement vector decreases until it reaches a minimum value and then increases again, but without divergence. In the (ICII) scenario, which corresponds to the case in which a body undergoes an explosion, we do not observe significant differences in the radial component of the displacement vector when compared to the (ICI) case. The radial component still grows up to a maximum value, whose magnitude decreases as the cloud of strings parameters increase, and then decreases until it vanishes. The main difference between the (ICI) and (ICII) cases in the radial sector is essentially the amplitude, since in the (ICII) case the separation reaches a much larger maximum value than in the (ICI) case. However, the angular component exhibits a more pronounced difference between the two scenarios. In the (ICII) condition, the initial angular separation is set to zero and then increases due to the relative velocity of separation. As a result, the angular component starts from zero, grows until it reaches a maximum value, then decreases to a minimum, and finally increases again, approaching a finite but nonzero value. In the Schwarzschild case, this value would vanish due to the infinite stretching experienced by the body. Therefore, we conclude that the cloud of strings can modify the usual stretching effect expected in gravitational systems. Although a compression effect becomes present, it remains hidden by the presence of the event horizon.

In the second part of our analysis, we focused on circular motion and first examined the asymptotic regime ($r_c \gg M$) of the tidal field. We found that even far from the central source, the influence of the string medium persists through the dependence on the parameters $l_s$ and $g_s$, affecting the stability of circular orbits. Physically, this means that the cloud of strings modifies not only the local curvature near the BH, but also the global tidal structure of the spacetime. We showed that, in the asymptotic regime $r \gg M$, the presence of the cloud of strings induces a tidal anisotropy, evidenced by the difference between the characteristic frequencies $\omega_{\text{eff}}$ and $\omega_{\theta}$. This result suggests that the cloud of strings may give rise to precession and resonance phenomena in the particle orbits even far away from the central source $M$.

Although the full analytical expressions of the tidal forces are too cumbersome to provide direct physical insight, their global behavior can still be clearly understood from the numerical plots in Figs.~\ref{fig:k1c} and ~\ref{fig:k2c}. These figures allow us to identify the main qualitative trends of the tidal field in the Letelier-Alencar geometry.

For a fixed circular orbit radius, increasing $g_s$ systematically weakens the radial tidal field. At the same time, a larger $g_s$ pushes the ISCO outward, so that stable circular orbits occur farther from the center, where the radial stretching is naturally smaller. Varying $l_s$ leads to a similar suppression for orbits at the same radius. However, because larger $l_s$ permits stable orbits much closer to the central region, the radial force can grow significantly in those inner zones.

The angular component exhibits a complementary behavior. For orbits at the same radius, increasing $g_s$ tends to strengthen the angular tidal force. Yet the outward shift of the ISCO for larger $g_s$ places the allowed circular orbits in regions where this component becomes weaker, so smaller $g_s$ values ultimately produce the highest angular intensities. The dependence on $l_s$ follows the same pattern as in the radial case: for a fixed radius, the tidal strength decreases with increasing $l_s$, but the smaller radii accessible for larger $l_s$ compensate for this, leading to stronger angular forces closer in.

Throughout this discussion, we refer to the absolute magnitude of the tidal forces. No sign reversals appear in the components, so the system never transitions from stretching to compression. Our analysis is restricted to stable circular orbits ($r_c \geq r_I$), since orbits inside the ISCO are unstable and do not represent physically meaningful configurations for sustained tidal interaction.

Overall, these results show that the parameters of the cloud of strings present in the Letelier-Alencar spacetime influence the tidal field in two distinct ways: by modifying the local curvature directly and by altering the location of the ISCO. Because of this dual effect, tidal observables, such as distortion limits for extended bodies or thresholds for tidal disruption, could, in principle, carry signatures of the underlying string distribution. This opens the possibility that precise tidal measurements around compact objects may help distinguish standard matter sources from more exotic configurations like those considered here.

Finally, our results open several directions for future work. A natural extension is to refine the analysis of tidal forces by including, for instance, the Roche limit to characterize the disruption of extended bodies in the vicinity of the Letelier-Alencar BH \cite{Junior:2026upy,Silva:2026koi}. Another interesting avenue is to generalize the present study to rotating configurations in order to assess how the angular momentum of the cloud of strings modifies the tidal force profile. In addition, the BH solution itself can be further explored from an observational perspective by computing its shadow and the deflection of light rays, thus enabling comparisons of the model with gravitational lensing and BH imaging data.

\section*{Acknowledgments}
\hspace{0.5cm} The authors thank the Conselho Nacional de Desenvolvimento Cient\'{i}fico e Tecnol\'ogico (CNPq). This study was financed in part by the Coordenação de Aperfeiçoamento de Pessoal de Nível Superior – Brasil (CAPES), Finance Code 001. The authors also acknowledge financial support from the  Spanish Grant PID2023-149560NB-C21, funded by MCIN/AEI /10.13039/501100011033. The paper is also supported by the Spanish project PID2024-157196NB-I00 funded by MICIU/AEI/10.13039/501100011033; the Q-CAYLE project, funded by the European Union-Next Generation UE/MICIU/Plan de Recuperacion, 
Transformacion y Resiliencia/Junta de Castilla y Le\'{o}n (PRTRC17.11) and the financial support by the Department of Education, Junta de Castilla y Le\'{o}n and FEDER Funds, 
Ref.~CLU-2023-1-05.
\bibliographystyle{apsrev4-1}
\bibliography{ref.bib}

\end{document}